\documentstyle[aps,eqsecnum,preprint,floats,epsf,psfig]{revtex}

\tighten

\begin{document}

\draft
\pagestyle{empty}

\preprint{
\noindent
\hfill
\begin{minipage}[t]{3in}
\begin{flushright}
LBNL--40822 \\
UCB--PTH--97/47\\
hep-ph/9709411 \\
September 1997
\end{flushright}
\end{minipage}
}

\title{Third Generation Familons, \\
$\bbox{B}$ Factories, and Neutrino Cosmology
}
\author{Jonathan L. Feng,$^{1,2}$ 
\thanks{Research Fellow, Miller Institute for Basic Research in
Science.} 
Takeo Moroi,$^1$\\
Hitoshi Murayama,$^{1,2}$
\thanks{Supported by the Alfred P. Sloan Foundation.}
and Erhard Schnapka$^1$
\thanks{Supported by the BASF--AG and the 
        Studienstiftung des deutschen Volkes.} }

\address{$^1$ Theoretical Physics Group \\
Ernest Orlando Lawrence Berkeley National Laboratory\\
University of California, Berkeley, California 94720 }

\address{$^2$ Department of Physics \\
University of California, Berkeley, California 94720 }

\maketitle

\begin{abstract}
We study the physics of spontaneously broken family symmetries acting
on the third generation.  Massless familons (or Majorons) $f$
associated with such broken symmetries are motivated especially by
cosmological scenarios with decaying tau neutrinos.  We first note
that, in marked contrast with the case for the first two generations,
constraints on third generation familon couplings are poor, and are,
in fact, non-existent at present in the hadronic sector. We derive new
bounds from $B^0$--$\bar{B}^0$ mixing, $B^0 \to l^+ l'^-$, $b\to
s\nu\bar{\nu}$, and astrophysics.  The resulting constraints on
familon decay constants are still much weaker than those for the first
and second generation.  We then discuss the promising prospects for
significant improvements from searches for $\tau\to l f$, $B\to (\pi,
K) f$, and $b\to (d,s) f$ with the current CLEO, ARGUS, and LEP data.
Finally, we note that future constraints from CLEO III and the $B$
factories will probe decay constants beyond $10^8$~GeV, well within
regions of parameter space favored by proposed scenarios in neutrino
cosmology.
\end{abstract}

\pagestyle{plain}

\section{Introduction}
\label{sec:introduction}

For over half a century, one of the major puzzles in particle physics
has been the question of why quark and lepton families replicate.
Although we have accumulated a wealth of data concerning the masses
and mixings of quarks and leptons, we still appear to be far from a
true understanding of family structure.  In the absence of a concrete
model to consider, it is natural to postulate the existence of some
family symmetry~\cite{Wilczek,Reiss,GNY} that plays a role in
determining the observed particle spectrum.  Once we consider such a
family symmetry, we face a plethora of options. The symmetry may be
(1) discrete,\footnote{There is a subtle distinction between global
and gauged discrete symmetries~\cite{KW}.  For this phenomenological
analysis, however, they are equivalent.} (2) continuous and local, or
(3) continuous and global.  Within each of these categories, one may
choose any of a number of symmetry groups, and the overall family
symmetry may even be a combination of the three possibilities.

Of course, any exact family symmetry of the underlying theory must be
spontaneously broken at some energy scale since we know that the quark
and lepton masses are very different from one family to the next.  For
option (1), spontaneously broken discrete symmetries, domain walls are
the only model-independent predictions, and these cannot be studied in
particle physics laboratories. In case (2), the masses of the family
gauge bosons of spontaneously broken local continuous symmetries can
be constrained, {\em e.g.}, from $K^0$-$\bar{K}^0$
mixing~\cite{Yanagida}.

From a phenomenological point of view, however, possibility (3) is
particularly enticing, as it implies the existence of massless
Nambu--Goldstone bosons, called ``familons,'' from the spontaneously
broken family symmetry.  This family symmetry may be either Abelian or
non-Abelian; Nambu--Goldstone bosons associated with the
spontaneous symmetry breaking of an Abelian lepton number symmetry are
often called ``Majorons.''\footnote{Majorons have been extensively
studied, and arise in a variety of models~\cite{Valle}, including, for
example, supersymmetric theories with spontaneous $R$-parity
breaking~\cite{Romao}.  In this paper, we study a number of probes,
many of which are applicable to both Abelian and non-Abelian
symmetries.  We use the generic name ``familon'' to denote the
associated Nambu--Goldstone bosons in either case.}  The existence of
new massless particles has many implications in particle physics,
astrophysics, and cosmology, and, as we will see, may be probed in a
wide variety of experiments.  Moreover, the couplings of familons at
low energies are determined by the non-linear realization of the
family symmetry. These couplings are, {\em e.g.}, of the form

\begin{equation}
\label{coupling}
\frac{1}{F} \partial_{\mu}  f^a \, \bar{\psi}_L^i 
\gamma^{\mu} T^a_{ij} \psi_L^j \ ,
\end{equation}
where $F$ is the family symmetry breaking scale, {\em i.e.}, the
familon decay constant, $ f^a$ are the familons, $T^a$ are the
generators of the broken symmetry, and the $\psi_L$ are fermion fields in
terms of which the flavor symmetry is defined.  The strength of the
familon coupling is therefore inversely proportional to $F$ and can be
constrained for a given family symmetry group in a model-independent
manner.

Familon couplings between the first and second generations have been
studied extensively and will be reviewed below.  In contrast, however,
couplings involving the third generation are largely unexplored,
although they may have rather rich phenomenological and cosmological
implications~\cite{bkk}.  Current constraints in the lepton sector are
relatively weak, with the best bounds coming from $\tau \to (e, \mu)
f$ bounds~\cite{albrecht}, and there are at present no corresponding
bounds reported in the hadronic sector (see, however,
Ref.~\cite{FY}). At the same time, it is a logical possibility that
the familon couples preferentially to the third generation, and models
have been proposed in which this is the case~\cite{bk}.  It is
therefore interesting to explore the possibilities for improving (or
setting) bounds on familon scales for the third generation, especially
in light of the upcoming $B$ physics experiments.

In this paper, we will study what we believe to be the most sensitive
probes of couplings of familons to the third generation, primarily to
$\tau$ leptons and $b$ quarks.  We show that dedicated analyses of
existent data from CLEO, ARGUS, and LEP could probe family symmetry
breaking scales up to $\sim 10^7$~GeV and may be significantly
improved at future $B$ factories.  Simply because this is largely
unexplored physics, there is a high discovery potential for familons
at these facilities.

Familon couplings to the third generation are also of interest from a
cosmological point of view.  The mass of the $\tau$ neutrino is still
allowed to be as large as 18.2~MeV experimentally~\cite{nutaumass}.  A
heavy $\tau$ neutrino has interesting consequences for both big-bang
nucleosynthesis
(BBN)~\cite{BBN&neut1,BBN&neut2,aph9612085,aph9705148} and
large scale structure
formation~\cite{oldLSS,prl72-3754,prd51-2669,aph9707143}, as will be
discussed in Sec.~\ref{sec:cosmology}. Since a heavy neutrino ($\geq
100 h^2$~eV, where $h$ is the expansion rate of the universe in units
of 100 km/sec/Mpc) must decay in order not to overclose the universe,
an invisible decay into a lighter neutrino and a massless boson, such
as a familon (or Majoron), is typically required.  (The three neutrino
mode is strongly disfavored and therefore the familon mode is most
preferred \cite{BBN&neut1}.)  There is therefore an interesting
interplay between experimental searches for familons and scenarios
requiring heavy neutrinos, and, as we will see, future collider
experiments and analyses may severely constrain a number of such
cosmologically motivated scenarios.

This paper is organized as follows.  We begin in
Sec.~\ref{sec:familons} with a discussion of familon interactions.  In
particular, we emphasize that the familon interactions of particles in
the same gauge multiplet are expected to be comparable.  In
Sec.~\ref{sec:collider} we consider constraints on familon
interactions that may be inferred from current experimental data,
concentrating on familon couplings to the third generation.  Current
bounds on third generation couplings from astrophysical considerations
are presented in Sec.~\ref{sec:astrophysics}.  We then describe some
promising prospects for detecting familons in $B$ physics at future
experiments in Sec.~\ref{sec:Bfactories}.  Finally, we note some of
the interesting cosmological implications in Sec.~\ref{sec:cosmology}
and give our conclusions in Sec.~\ref{sec:conclusions}.

\section{Familon Interactions}
\label{sec:familons}

The standard model contains 15 particle states in each of the 3
generations.  These states are distinguished by the
SU(3)$_C\times$SU(2)$_L\times$U(1)$_Y$ gauge interactions, which
divide each generation into 5 multiplets: $Q$, $U$, $D$, $L$, and
$E$. The gauge interactions therefore break the flavor symmetry group
from U(45) to U(3)$^5$. In the standard model, the flavor group
U(3)$^5$ is broken explicitly to $\text{U(1)}_B \times \text{U(1)}_L$
by Yukawa couplings. However, in extensions of the standard model in
which one hopes to gain some understanding of the pattern of fermion
masses and mixings, some subgroup of the flavor group may be an exact
symmetry of the Lagrangian that is broken spontaneously by the vacuum,
and it is this possibility we consider here.

The massless Nambu-Goldstone bosons of the spontaneously broken flavor
symmetry, familons~\cite{Wilczek,Reiss,GNY}, have interactions given
by the couplings 

\begin{equation}
{\cal L}_{ f} = \frac{1}{F} \partial_{\mu} 
 f^a \, J^{\mu a} \ ,
\end{equation}
where $ f^a$ are the familon fields, and $J^{\mu a}$ are flavor
currents.\footnote{Throughout this study, we will assume that no
additional light degrees of freedom are introduced by other new
physics.}  The interactions are suppressed by $F$, the scale at which
the flavor symmetry is spontaneously broken.  Note that familons are
derivatively coupled,\footnote{If the flavor symmetry is anomalous,
familons may also have non-derivative, flavor-diagonal couplings.  We
will not consider such couplings here.} and so do not mediate
long-range ($\sim r^{-2}$) forces. The most general current $J^{\mu
a}$ composed of two fermion fields takes either the form

\begin{equation}
\label{AV}
J^{\mu a} = \bar{\psi}_i \gamma^{\mu} (g_V + g_A \gamma_5 ) T_{ij}^a
\psi_j
\end{equation}
or
\begin{equation}
J^{\mu a} = \bar{\psi}_i \gamma^{\mu} (g_L P_L + g_R P_R ) 
T_{ij}^a \psi_j \ ,
\end{equation}
where $P_{L,R}$ are the projection operators $\frac{1}{2} (1\pm
\gamma_5)$, $i$ and $j$ are generational indices, and $T_{ij}^a$ are
the spontaneously broken generators of the family symmetry.  The
fields $\psi_i$ and $\psi_j$ are fermion mass eigenstates, which we
assume here to be also flavor eigenstates.  (The more general case is
described below.)  Using the form of the current given in
Eq.~(\ref{AV}), the familon interaction may be written as

\begin{eqnarray}
\label{massform}
{\cal L}_{ f} &=& \frac{1}{F} \partial_{\mu}  f^a \,
\bar{\psi}_i \gamma^{\mu} (g_V + g_A \gamma_5 ) T_{ij}^a \psi_j 
\nonumber \\
&=& -\frac{i}{F}  f^a \bar{\psi}_i \left[ g_V (m_i - m_j) 
+ g_A (m_i+m_j) \gamma_5 \right] T_{ij}^a \psi_j \ ,
\label{famint}
\end{eqnarray}
where in the last step we have integrated by parts and then
substituted the equations of motion. The second line of
Eq.~(\ref{famint}) is of course only valid for on-shell fermions such
as external leptons, whereas in hadronic matrix elements and processes
including off-shell fermions, the derivative coupling of the first
line must be used.

We see that familons may mediate or be produced in family-changing
processes.  They may also couple to identical fermions
$\psi_i=\psi_j$, but only through axial couplings.  What processes are
mediated by familons depends on the particular family symmetry group
that is broken.  For example, for O($N$) groups, the generators
$T_{ij}$ are anti-symmetric, and so do not generate flavor-diagonal
interactions.  However, they do generate interactions like $ f
\bar{\psi}_i \gamma_5 \psi_j - f \bar{\psi}_j \gamma_5 \psi_i$, where 
we have considered axial vector current interactions as an example.
Familons from O($N$) groups may therefore mediate neutral meson
mixing, which we will consider in Sec.~\ref{sec:mesons}.  The
situation is reversed for SU($N$) groups.  Here, flavor-diagonal
couplings exist.  However, if we consider any SU(2) subgroup and form
the complex familon $\tilde{f} = f^1 + i f^2$, the off-diagonal
interactions are given by $\tilde{f} \bar{\psi}_i \gamma_5 \psi_j +
\tilde{f}^{*} \bar{\psi}_j \gamma_5 \psi_i$, and we see that $\tilde{f}$
exchange cannot induce neutral meson mixing.

Up to this point, we have ignored possible mass mixing effects. In
general, if the flavor eigenstates $\psi'$ are related to the mass
eigenstates $\psi$ by

\begin{equation}
\psi' = U_\psi \psi \ ,
\end{equation}
where $U_\psi$ is a $3\times 3$ unitary mixing matrix, the familon
interactions are given by

\begin{eqnarray}
{\cal L}_{ f} &=& \frac{1}{F} \partial_{\mu}  f^a \, 
\bar{\psi}'_i \gamma^{\mu} (g_V + g_A \gamma_5 ) T_{ij}^a \psi'_j 
\nonumber \\
&=& \frac{1}{F} \partial_{\mu}  f^a \,
\bar{\psi}_i \gamma^{\mu} (g_V + g_A \gamma_5 ) T_{\psi ij}^a \psi_j \ ,
\end{eqnarray}
where $T_{\psi}^a = U^{\dagger}_\psi T^a U_\psi$.  Mass mixings may
therefore generate flavor-diagonal interactions from flavor
off-diagonal interactions, and vice versa.  For example, in the case
of an Abelian U(1) symmetry, mass mixing effects may generate
flavor-changing interactions. They may also extend non-maximal family
symmetries to couplings involving all three generations; for example,
a U(2) symmetry between the first and second families, may, after
rotation to mass eigenstates, result in familon interactions involving
the third generation.

While the phenomenology of familons varies from group to group, it is
important to note that gauge symmetry relates the familon interactions
of particles in the same gauge multiplet.  As an example, let us
consider a spontaneously broken lepton flavor symmetry.  The familon
interaction is then given by

\begin{equation}
\frac{g_L}{F} \partial_{\mu}  f^a \,  
\bar{L}'_i \gamma^{\mu} T_{ij}^a L'_j \ , 
\end{equation}
where the SU(2) lepton doublets $L'_i = (\nu'_i, l'_{i})$ are in the
flavor eigenstate basis. This interaction therefore 
generates familon interactions
for both the charged leptons and neutrinos.  In the presence of
neutrino masses, the flavor eigenstates may not correspond to mass
eigenstates.  The familon interactions in the mass basis are then

\begin{equation}
\frac{g_L}{F} \partial_{\mu}  f^a \,  
\bar{\nu}_i \gamma^{\mu} T_{\nu ij}^a {\nu}_j 
+ \frac{g_L}{F} \partial_{\mu}  f^a \,  
\bar{l}_i \gamma^{\mu} T_{l ij}^a l_j \ , 
\label{massbasis}
\end{equation}
where $T_{\nu}^a = U_{\nu}^{\dagger} T^a U_{\nu}$, and we have defined
$V = U_{\nu}^{\dagger} U_l$ and $T_l^a = V^{\dagger} T_{\nu}^a V$.
$T_l^a$ and $T_{\nu}^a$ are therefore related by a similarity
transformation, and in the presence of mass mixing, the couplings of
the interactions of $\partial_{\mu}  f^a \, \bar{\nu}_i
\gamma^{\mu}\nu_j$ and $\partial_{\mu}  f^a \, \bar{l}_i \gamma^{\mu} 
l_j$ are not necessarily identical. However, in the absence of
fine-tuning, we expect these couplings to be of the same magnitude.
Bounds on one familon interaction may thus be considered to imply
comparable bounds on the other interactions linked by gauge symmetry.

Because the familon interactions of particles in the same gauge
multiplet are comparable in the absence of fine-tuning, there are many
more relations in theories with enlarged gauge groups.  For example,
for SU(5) grand unified theories (GUTs), the particles $d_R, \nu, e_L
\subset \overline{\bbox{5}}$ are expected to have comparable familon
interactions, as are the particles $u_L, d_L, u_R, e_R \subset
\bbox{10}$.  A particularly relevant example for our study below is
that, in the GUT framework, bounds on familon decays of $B$ mesons
imply bounds on familon decays of tau neutrinos in the absence of
fine-tuning.

Flavor mixing effects also induce familon couplings of fields with
different generational indices.  In the quark sector, for example,
substituting the quark doublet $Q'_i = (u'_i, d'_i)$ for $L'_i$ in the
discussion above, Eq.~(\ref{massbasis}) becomes

\begin{equation}
\frac{g_L}{F} \partial_{\mu}  f^a \,  
\bar{u}_i \gamma^{\mu} T_{u ij}^a {u}_j 
+ \frac{g_L}{F} \partial_{\mu}  f^a \,  
\bar{d}_i \gamma^{\mu} T_{d ij}^a d_j \ , 
\label{quarkmassbasis}
\end{equation}
where $T_u^a$ and $T_d^a$ are related by the CKM matrix through

\begin{equation}
T_d^a = V^{\dagger}_{\text{CKM}} T_{u}^a V_{\text{CKM}} \ .
\label{CKMrelation}
\end{equation}
We see that in general, couplings to all generations are induced by
flavor mixings.  For example, a familon with flavor-diagonal coupling
to $\bar{t}t$ in the up sector couples not only to $\bar{b} b$, but
also to, for example, $\bar{b} s$ and $\bar{d}d$.  The induced
couplings to first and second generation quarks in this case are
CKM-suppressed, but may still lead to significant bounds when, as is
often the case, these induced couplings are much more strongly
constrained.  We will consider the constraints on mixing-induced
couplings from $K$ decays in Sec.~\ref{sec:decays} and from supernova
cooling in Sec.~\ref{sec:mixing}.

Finally, note in Eq.~(\ref{AV}) that the strength of the interaction
depends not only on $F$, but also on $T_{ij}^a$ and the couplings
$g_{V,A}$. In the following sections, we will present a variety of
bounds on combinations of these couplings, and it is important that we
define our conventions and normalizations.  We will always define our
interaction as

\begin{equation}
\frac{1}{F} \partial_{\mu}  f \, \bar{\psi}_i \gamma^{\mu} 
(g_V^{ij} + g_A^{ij} \gamma_5 ) \psi_j \ ,
\label{conven}
\end{equation}
and similarly for $g_L^{ij}$ and $g_R^{ij}$; the superscripts of the
couplings will often be omitted when they are obvious from the
context.  In presenting our bounds, it will be convenient to define

\begin{equation}
F_{ij}^I \equiv F / g^{ij}_I \ ,
\end{equation}
where $I=V,A,L,R$. In addition, as many of our bounds are to a good
approximation independent of the chirality of the interaction and so
only dependent on the combination $g_V^{ij\,2} + g_A^{ij\,2}$, we
define

\begin{equation}
F_{ij} \equiv \frac{F}{\sqrt{g^{ij\,2}_V + 
g^{ij\,2}_A }} \ .
\label{parinv}
\end{equation}

\section{Bounds from accelerator data}
\label{sec:collider}

As described in the previous section, familons may take part in
flavor-changing processes, and bounds on such processes lead to lower
bounds on the familon energy scale.  For familons mediating
transitions between the first and second generation, such bounds are
rather stringent. In contrast, similar bounds involving the third
generation are much weaker, with the previously reported constraints
limited only to bounds from rare $\tau$ decays.  We are thus motivated
to focus on the third generation.  In Sec.~\ref{sec:decays}, we begin
by reviewing and contrasting such bounds, and then discuss the
implications of flavor eigenstate mixings.  We then go on to derive
new bounds from a variety of processes. In Sec.~\ref{sec:mesons} we
consider familon-mediated processes such as neutral meson mixing and
rare leptonic decays of mesons. Finally, in Sec.~\ref{sec:LEP} we
consider possible analyses at LEP and extrapolate a preliminary ALEPH
bound on $b \to s \nu \bar{\nu}$ to a bound on $b\to s f$.

\subsection{Decays to familons}
\label{sec:decays}

We begin by considering bounds from decays of mesons and leptons to
familons.  Normalizing the relevant familon scale according to
Eq.~(\ref{conven}), we find

\begin{equation}
  \Gamma(K \rightarrow \pi  f) 
  = \frac{1}{16\pi} \frac{m_K^3}{F^2} g_V^2 \beta^3 |F_1(0)|^2 \ ,
\end{equation}
where $\beta = 1 - m_\pi^2/m_K^2$. In the limit of exact flavor SU(3)
symmetry, the form factor $\langle \pi^+ (p')| \bar{s}\gamma^\mu d|K^+
(p)\rangle = F_1 (q^2) (p + p')^\mu$ at zero momentum transfer has a
fixed normalization, $F_1(0)=1$. For leptonic decays $l_i \to l_j
 f$, the exact tree-level partial decay width in the limit of
massless $l_j$ is given by
        
\begin{eqnarray}
\Gamma(l_i^- \to l_j^-  f) 
= \frac{1}{16 \pi}\frac{m_{l_i}^3}{F^2} (g_V^2+g_A^2) \beta^3 \ ,
\label{hier}
\end{eqnarray}
where here $\beta = 1 - m_{l_j}^2/m_{l_i}^2$.  

The strongest bound on any flavor scale is derived from the constraint
on exotic $K$ decay. Using the above expressions, the experimental
result $B(K^+ \to \pi^+  f) < 3.0 \times 10^{-10} \text{ (90\%
CL)}$~\cite{hex9708031} leads to the bound

\begin{eqnarray}
 F^V_{sd} > 3.4 \times 10^{11}\text{ GeV} \ .
\label{kpifam}
\end{eqnarray}
Note that the limit on $B(K^+ \to \pi^+  f)$ bounds only the
vectorial familon coupling; the axial coupling is unconstrained.  For
the leptonic sector, Jodidio {\em et al.} report the constraint
$B(\mu^+ \to e^+  f) < 2.6 \times 10^{-6} \text{ (90\%
CL)}$~\cite{jodidio}, which they obtain under the assumption of a
vector-like familon coupling. This can be converted into the bound

\begin{eqnarray}
 F^V_{\mu e} > 5.5 \times 10^{9}\text{ GeV} \ .
\end{eqnarray}
For familon interactions of arbitrary chirality, the slightly weaker
constraint 

\begin{eqnarray} 
 F_{\mu e}>3.1\times 10^9 \text{ GeV}
\end{eqnarray}
may be obtained from the bound $B(\mu^+ \to e^+ \gamma  f) < 1.1
\times 10^{-9}$ (90\% CL)~\cite{bolton}.

We now compare these bounds to those available in the third
generation.  The ARGUS collaboration~\cite{albrecht} has bounded the
branching fractions of $\tau$ decays into light bosons and found the
limits $B(\tau^- \to \mu^-  f) < 4.6 \times 10^{-3}\text{ (95\%
CL)}$ and $B(\tau^- \to e^-  f) < 2.6 \times 10^{-3}\text{ (95\%
CL)}$. These imply the following constraints on the flavor scale:

\begin{eqnarray}
   F_{\tau \mu} & > & 3.2 \times 10^6\text{ GeV} \\
   F_{\tau e}   & > & 4.4 \times 10^6\text{ GeV} \ . 
\end{eqnarray}
 
We see that the bounds on flavor scales in the leptonic sector are
significantly less stringent for third generation couplings than for
those involving only the first two.  The discrepancy is even more
pronounced in the hadronic sector, where there are as yet no reported
bounds on flavor scales from $B$ decays.  

It is also worth noting, however, that strong bounds on a particular
flavor scale, such as the one on $F_{sd}^V$, may imply significant
bounds on other flavor scales as well. These bounds are induced by the
flavor-mixing effects discussed in Sec.~\ref{sec:familons} and are
thus model-dependent.  As an example let us now assume that flavor and
mass eigenstates coincide for up-type quarks. A given familon coupling
in the up sector requires, by gauge invariance, a corresponding
coupling in the down sector. For example, from
Eqs.~(\ref{quarkmassbasis}) and (\ref{CKMrelation}) we see that the
coupling $\partial_\mu f \, \bar{t}\gamma^\mu P_L c /F^L_{tc}$ induces
the coupling $V_{ts}^* V_{cd} \partial_\mu f \, \bar{s}\gamma^\mu P_L
d /F^L_{tc}$, which mediates the rare decay $K^+ \rightarrow \pi^+ f$.
Assuming complex familons, the Hermitian conjugate coupling gives a
similar contribution $\propto V_{cs}^* V_{td}$ to the decay into the
complex conjugate familon.  Summing both decay widths and comparing to
the bound on $F_{sd}^V$ in Eq.~(\ref{kpifam}), one can derive the
mixing induced bound

\begin{eqnarray}
 F^L_{tc} > 2.2 \times 10^{9}\text{ GeV} \ .
\end{eqnarray}
Under similar assumptions, we find $F^L_{tu} > 6.6 \times 10^{9}\text{
GeV}$. Note, however, that such bounds do not apply if the mass and
flavor bases are aligned in the down sector~\cite{ns} or if the
couplings are purely axial.

\subsection{Familon-mediated processes}
\label{sec:mesons}

In this section we derive new constraints on the scale of spontaneous
flavor symmetry breaking by considering non-standard familon
contributions to neutral meson mixing and existing bounds on rare
leptonic decays such as $B^0 \to \tau e$.

A familon contribution to neutral meson mixing requires a real flavor
group to be spontaneously broken in the corresponding sector, such
that the same real familon scalar field couples to the quark current
and its Hermitian conjugate. For concreteness, let us consider the
$B^0-\bar{B}^0$ system; similar formulae hold (at least approximately)
for other neutral meson systems. Assuming the general coupling
structure
\begin{eqnarray}
\frac{i}{F} \partial_{\mu}  f  \left[
 \bar{d}\gamma^{\mu} (g_V + g_A \gamma_5) b 
  - \bar{b}\gamma^\mu (g_V + g_A \gamma_5) d \right] \ ,
\label{mesonmixing}
\end{eqnarray}
we find a familon contribution to the mass splitting of 
\begin{eqnarray}
\Delta m^{( f)}_{B^0} \equiv \left|m_{B^0} - m_{\bar{B}^0}\right| 
\approx \frac{5}{6} \frac{f_{B^0}^2 g_A^2 m_{B^0}}{F^2} \ .
\label{deltamsq}      
\end{eqnarray}

Eq.~(\ref{deltamsq}) may be derived by taking the matrix element of
the non-local operator
\begin{eqnarray}
\frac{1}{2!}\frac{1}{F^2} \bar{d}_{\alpha} \gamma^{\mu} 
(g_V + g_A \gamma_5) b_{\beta} \frac{q_\mu q_\nu}{q^2} 
 \bar{d}_{\gamma} \gamma^{\nu} (g_V + g_A \gamma_5) b_{\delta}
\label{oper}
\end{eqnarray}
between $B^0$ and $\bar{B}^0$ states and using the definition of the
pseudoscalar decay constant, $\langle 0| \bar{b}\gamma^\mu \gamma_5 d
(0)|B^0 (p) \rangle = i f_{B^0} p^{\mu}$. The subscripts $\alpha$,
$\beta$, $\gamma$, and $\delta$ in Eq.~(\ref{oper}) are color indices.
Between two color singlet states, there are two contributions.  The
first one arises from $\alpha=\beta$ and $\gamma=\delta$ with a
familon in the $s$-channel.  In this case, the momentum transfer
through the familon propagator is $q^2=m_{B^0}^2$, and after a vacuum
insertion, it is easy to verify that this contribution is as in
Eq.~(\ref{deltamsq}), but without the factor of $5/6$.  However, there
is also a $t$-channel contribution from $\alpha = \delta$ and
$\beta=\gamma$, which may be evaluated by a Fierz transformation and
then a vacuum insertion as before. For a heavy--light system like the
$B^0$ meson, one may assume the free-quark picture, in which the
momentum transfer is governed by the energy of the ``static'' $b$
quark $q^0 \approx m_b \approx m_{B^0}$, and, in the numerator, the
derivative acting on the quark current gives again a factor of
$m_b$. Using $\langle 0|\bar{b}\gamma_5 d (0)|B^0 (p) \rangle \approx
i f_{B^0} m_{B^0}$ and including the relative color factor of $1/3$,
one can estimate the $t$-channel contribution to be $-1/6$ times the
$s$-channel contribution, which leads to Eq.~(\ref{deltamsq}).

Our result should be fairly reliable for the $B^0$ meson. For $D^0$
and $K^0$ mesons, the evaluation of the $t$-channel momentum transfer
is more ambiguous. However, because this contribution is suppressed
relative to the $s$-channel part, we expect the result of
Eq.~(\ref{deltamsq}) to be reasonably accurate in these cases as well.
We also note that a vector-like familon interaction does not
contribute to the mass splitting, at least in the heavy quark
approximation $m_b \approx m_{B^0}$. Although one might expect a
vector contribution to appear in the $t$-channel contribution after
the Fierz rearrangement, one finds that the term proportional to $g_V$
contains axial vector and pseudoscalar contributions of equal
magnitude but opposite sign.

The constraint on the flavor scale $F$ results in principle from the
requirement that the combined standard model and familon contributions
do not exceed the measured value. However, when considering
nonstandard contributions, it is also uncertain what one should take
as the standard model contribution. For example, the reported
value~\cite{PDG} for $|V^*_{tb}V_{td}|$ is derived from
$B^0-\bar{B}^0$ mixing under the assumption that the standard model
gives the only contribution. As a conservative bound, we simply
compare the familon contributions directly to the corresponding
measured values.  The results are summarized in
Table~\ref{tab:mixing}.  We take the decay constants to be $f_{B^0}
\approx 175$~MeV and $f_{D^0} \approx 205$~MeV from recent lattice
results~\cite{lattice96}, and $f_{K^0} \approx f_{K^+}
\approx 160$~MeV~\cite{PDG}.  Since we use the measured
mass splitting (not its error), the bounds from $B^0$ and $K^0$ will
only improve when the size of the standard model contribution can be
quantified independently. For the $D^0$, where only the upper bound on
the mass splitting is known, future experiments will improve the
bound.

\begin{table}
\caption{Bounds on the flavor scale from contributions to neutral 
meson mixing from familon exchange as given in
Eq.~(\protect\ref{deltamsq}).  Note that these limits do not apply to
vector-like couplings, and that this process requires a real flavor
group so that a real familon scalar field couples to a current
operator and its Hermitian conjugate, as in
Eq.~(\protect\ref{mesonmixing}).}
\begin{tabular}{rrr} 
                &  $\Delta m_{\text{exp}}$\hspace{3em}                     
& Bound\hspace{10mm} \\ \hline
$B^0-\bar{B}^0$ &  $0.5 \times 10^{12}\, \hbar s^{-1}\;$~\cite{PDG}   
& $ F^A_{bd} > 6.4 \times  10^5$ GeV  \\ 
$D^0-\bar{D}^0$ & $ <21 \times 10^{10}\, \hbar s^{-1}\;$~\cite{anjos} 
& $ F^A_{cu} > 6.9 \times  10^5$ GeV  \\
$K^0-\bar{K}^0$ & $0.53 \times 10^{10}\, \hbar s^{-1}\;$~\cite{PDG}   
& $ F^A_{sd} > 1.7 \times  10^6$ GeV \\ 
\end{tabular}
\label{tab:mixing}
\end{table}

We next consider rare leptonic decays of neutral mesons, mediated by
familon exchange. Such decays are possible if the same familon couples
to both quarks and leptons. This is guaranteed in grand unified
scenarios, where quarks and leptons are in the same gauge multiplet.
In general the relevant interaction can be written in terms of
effective vector and axial vector couplings that parametrize the
familon couplings and mixing angles of a particular model. For
example, the process $B^0 \to \tau^+ e^-$ can be mediated by the
interaction Lagrangian
\begin{eqnarray}
\frac{1}{F} \partial_{\mu}  f \left[
 \bar{b}\gamma^{\mu} (g^{bd}_V+g^{bd}_A \gamma_5 ) d + 
 \bar{\tau}\gamma^{\mu} (g^{\tau e}_V+g^{\tau e}_A \gamma_5 ) e 
  \right] + \text{ h.c.}  
\label{rareB}
\end{eqnarray}
Note that the constants $g_V$ and $g_A$ may be different in the
hadronic and leptonic sectors. Also, even if familon couplings always
include third generation flavor eigenstates, mixing effects may induce
transitions like $B^0 \to \mu^+ e^-$.

With the interaction defined in Eq.~(\ref{rareB}) one obtains a width
of
\begin{eqnarray}
\Gamma (B^0 \to \tau^+ e^-) \approx 
\frac{1}{8\pi} \frac{f_{B^0}^2 g_A^{bd \,2} m_{B^0} m_\tau^2 }{F^4} 
\left[  \left( g^{\tau e\,2}_V + g^{\tau e\,2}_A \right) \beta^2 
  - 2\frac{m_e}{m_\tau} 
        \left( g^{\tau e\,2}_V - g^{\tau e\,2}_A \right) \beta
     \right] \ ,
\label{Gammarare}
\end{eqnarray}
where $\beta = 1 - m_{\tau}^2 / m_{B^0}^2$, and we have displayed the
leading $g_V^2 - g_A^2$ piece.  In the limit where the lighter lepton
is massless, the result is independent of the chirality of the
interaction and depends only on the combination of lepton couplings
$g^2_V + g^2_A$. Expressions for other similar processes are obtained
by replacing the coupling constants $g_V, g_A$ and the meson and
lepton masses accordingly. Limits on the flavor scales from current
experimental bounds on rare leptonic decays are given in
Table~\ref{tab:decay}.  

The bounds of Tables~\ref{tab:mixing} and \ref{tab:decay} are
significantly weaker than those presented in Sec.~\ref{sec:decays}.
This is especially true in Table~\ref{tab:decay}, as rare leptonic
meson decays are dependent on the flavor scale to the fourth
power. However, such processes set bounds on third generation hadronic
familon couplings, which were previously unconstrained.  It is also
important to note that the bounds on familon couplings to the first
two generations are also interesting, as they constrain axial
couplings, whereas the bound from $K$ decay reviewed in the previous
section bounds only vector-like couplings.

\begin{table}
\caption{Limits on flavor scales and couplings for some rare meson
decays. The branching ratio bounds are on the sum of the two charge
states, assuming real familon scalars that mediate both decay modes. 
If familons mediate only one decay mode, the quoted bounds on 
$F$ are weakened by a factor of $2^{1/4}$. 
In calculating these bounds, we neglect small corrections 
from the lighter lepton mass.}
\begin{tabular}{rrr} 
         & Branching Ratio Upper Bound&  Bound\hspace{10mm} \\ 
\hline
$B^0 \to\tau^{\pm} e^{\mp}$   & $ 5.3 \times 10^{-4} $~\cite{ammar} 
         & $(F^A_{bd} F_{\tau e})^{1/2} > 3.5 \times  10^3$ GeV \\
$B^0 \to \tau^{\pm}\mu^{\mp}$ & $ 8.3 \times 10^{-4}$~\cite{ammar}  
         & $(F^A_{bd} F_{\tau \mu})^{1/2} > 3.1 \times 10^3$ GeV \\
$B^0 \to \mu^{\pm} e^{\mp} $  & $ 5.9 \times 10^{-6}   $~\cite{ammar}  
         & $(F^A_{bd} F_{\mu e})^{1/2} > 2.8 \times 10^3$ GeV \\
$D^0 \to \mu^{\pm} e^{\mp} $  & $ 1.9 \times 10^{-5} $~\cite{freyb} 
         & $(F^A_{cu} F_{\mu e})^{1/2} > 1.2 \times 10^3$ GeV \\
$K_L^0 \to \mu^{\pm} e^{\mp}$ & $3.3 \times 10^{-11} $~\cite{arisaka} 
         & $(F^A_{sd} F_{\mu e})^{1/2} > 3.8 \times 10^5$ GeV \\ 
\end{tabular}
\label{tab:decay}
\end{table}

\subsection{Constraints from LEP}
\label{sec:LEP}

Currently, there are no reported experimental bounds on decays $b\to
(s,d)  f$.  One can, however, infer a constraint from ALEPH's
preliminary bound on $b\to s\nu\bar{\nu}$~\cite{ALEPHmissing}.  By
searching for events with large missing energy, they placed the
constraint $B(b \to s\nu\bar{\nu}) < 7.7 \times 10^{-4}$ (90\% CL).
One can rescale this constraint to obtain an upper bound on $B(b\to
s f)$.

The analysis for $b\to s\bar{\nu}\nu$ relies on the $E_{\text{miss}}$
distribution~\cite{ALEPHtau}, where $E_{\text{miss}}$ is defined by
\begin{equation}
        E_{\text{miss}} = E_{\text{beam}} + E_{\text{corr}} -
        E_{\text{vis}}
\end{equation}
in each hemisphere of $b$-tagged events.  Here, $E_{\text{beam}}$ is
half of the center-of-momentum energy, $E_{\text{corr}} =
(M_{\text{same}}^{2} - M_{\text{opp}}^{2})/4E_{\text{beam}}$, where
$M_{\text{same}}$ and $M_{\text{opp}}$ are the visible invariant
masses in the same and opposite hemispheres, respectively, and
$E_{\text{vis}}$ is the total visible energy in the hemisphere.
$E_{\text{corr}}$ improves the estimate of the actual missing
energy in the hemisphere by correcting for the fact that the
hemisphere with larger invariant mass typically has higher energy.

The backgrounds from $b\to l\nu X$ and $c\to l\nu X$ are suppressed by
rejecting events with identified $e^{\pm}$ or $\mu^{\pm}$ in the
relevant hemisphere.  Up to this point, we do not expect significant
differences in efficiencies between the $b\to s\nu\bar{\nu}$ mode and
the $b\to s f$ mode.  They then required $35\text{ GeV}
<E_{\text{miss}}<45\text{ GeV}$.  The efficiencies for this
requirement obviously differ between the two decay modes, since the
mode $b\to s\nu\bar{\nu}$ has two missing neutrinos, resulting in a
harder $E_{\text{miss}}$ spectrum than that of the $b\to s f$ mode.
The $E_{\text{miss}}$ spectrum of both modes may be calculated by
convoluting the theoretical missing energy distribution in three-body
($s\nu\bar{\nu}$) and two-body ($s f$) decays with the measured $b$
fragmentation function~\cite{ALEPHxb}.  We find that the ratio of
efficiencies is 0.43 with little dependence on the details of the
fragmentation function.  By scaling the reported $B(b \to
s\nu\bar{\nu})$ upper bound by this factor, we find

\begin{equation}
        B(b \to s f) < 1.8 \times 10^{-3} \ .
\label{bsbound}
\end{equation}
Using the expression of Eq.~(\ref{hier}) with the substitution of
$m_{B^0} \approx m_b$ for $m_{l_i}$, this corresponds to a limit on
the flavor scale of

\begin{eqnarray}
    F_{bs} > 6.1 \times 10^7 \text{ GeV} \ .
\label{bsFbound}
\end{eqnarray}
Note that this analysis does not require an energetic strange
particle, and so the constraint of Eq.~(\ref{bsbound}) is actually on
the sum $B(b \to s\, f) + B(b \to d\, f)$. Thus, for $F_{bs}
\approx F_{bd}$, the bound on the flavor scale given in
Eq.~(\ref{bsFbound}) improves by a factor $\sqrt{2}$.  The bound of
Eq.~(\ref{bsFbound}) is enhanced by the fact that the SM decay width
is greatly suppressed by $V_{cb}$, which increases the sensitivity of
$b$ decays to small exotic decay widths.

\section{Bounds from Astrophysics}
\label{sec:astrophysics}

In this section, we discuss constraints on third generation familon
couplings from astrophysics.  We begin in Sec.~\ref{sec:tree} with
constraints on direct (tree-level) couplings.  Second and third
generation particles are absent in almost all astrophysical objects.
The exception is supernovae, where all three neutrino species are
thermalized in the core. We therefore consider what bounds on familon
couplings to $\tau$ neutrinos may be obtained by supernova
observations.  Couplings of familons to the third generation may also
radiatively induce couplings to first generation particles.  Although
such induced couplings are suppressed by loop factors, they are so
stringently bounded by constraints from supernovae, white dwarfs, and
red giants that interesting bounds also result.  These are studied in
Sec.~\ref{sec:loop}.  Finally, mixings of flavor eigenstates may also
induce couplings of familons to the first generation; such effects are
discussed in Sec.~\ref{sec:mixing}. It is important to note that,
while the bounds derived in this section are rather strong in certain
cases, they are also typically more model-dependent than, for example,
the accelerator bounds of the previous section.  We therefore specify
the necessary conditions for each bound in detail in each case.

\subsection{Bounds from direct couplings}
\label{sec:tree}

In 1987, the Kamiokande group and the IMB group independently detected
neutrinos emitted from supernova SN~1987A. They observed that the
neutrino pulse lasted for a few seconds.  Furthermore, their results
indicate that neutrinos carried off about $10^{53}$~erg from the
supernova.  The observed duration time and neutrino flux can be well
explained by the generally accepted theory of core collapse, and the
observations confirmed the idea that most of the released energy in
the cooling process is carried off by neutrinos.  Exotic light
particles, such as familons, may affect the agreement of theory and
observation, since they can also carry off a significant energy
fraction.  The core of the supernova is hot ($T\sim 30\text{ MeV}$)
and dense, and so neutrinos are thermalized in the core and can be a
source of familon emission. If the energy fraction carried away by
familons is substantial, the duration time of the neutrino pulse
becomes much shorter than the observed value. In order not to affect
the standard cooling process, the familon luminosity $Q_f$ must be
smaller than the neutrino luminosity, {\em i.e.}, less than $\sim
10^{53}$~erg/sec.

This constraint can be satisfied in two different regimes of the
familon coupling strength. For sufficiently high flavor scales $F$,
the familon interaction is weak enough that familons are rarely
produced and the familon luminosity $Q_f$ is suppressed. On the other
hand, for sufficiently low flavor scales, although familons are
readily produced, they interact so strongly that they become
thermalized and trapped in the core as well, thus decreasing the
familon luminosity.  Therefore, there are two parameter regions
consistent with observations, high and low $F$, with an excluded
region in the middle.

These types of constraints have been discussed by Choi and
Santamaria~\cite{PRD42-293} in the context of a Majoron model.  We
modify their discussions slightly for the familon case.  To simplify
the analysis, we will look at two extreme scenarios.  First we
consider a diagonal familon coupling to $\nu_\tau$, as in the case of
an Abelian family symmetry, and second we analyze a purely
off-diagonal coupling, as in the case of an O(2) family
symmetry.\footnote{Throughout our discussions, we assume that
neutrinos are Majorana particles.  Observations of supernova SN1987A
imply that Dirac neutrinos must be lighter than
3~keV~\cite{PLB317-119} or heavier than 31~MeV~\cite{PLB242-77}.  As
we will discuss in Sec.~\ref{sec:cosmology}, most of the interesting
mass range from a cosmological point of view is therefore excluded.}
For a general family symmetry, one expects familons with both diagonal
and off-diagonal couplings; a generalization to such cases is
straightforward.  In this subsection, we also neglect possible
mismatches between the flavor and mass eigenstates, and assume that
the relative angles relating the two are small.  
Such mismatches will be discussed in
Sec.~\ref{sec:mixing}.  Finally, we assume that $m_{\nu_e}$,
$m_{\nu_\mu}$ are negligible compared to $m_{\nu_\tau}$, as suggested
from laboratory constraints as well as the corresponding masses of the
charged leptons.

\subsubsection*{Familon with diagonal coupling}
Here we consider a purely diagonal familon coupling to $\nu_\tau$,
such as in models with a U(1) family symmetry acting on the
third-generation lepton doublet $(\nu_\tau, \tau_L)$.  The relevant
interaction is given by
\begin{equation}
  {\cal L}_{ f} = \frac{1}{F} g_L^{\nu_\tau \nu_\tau}
  \partial_\mu f \, \bar{\nu}_\tau\gamma^\mu P_L \nu_\tau \ .
\label{L_on}
\end{equation}

Let us first consider the case where the familon can freely escape the
core of the supernova. Based on the interaction given in
Eq.~(\ref{L_on}), potentially significant processes of familon
production are the neutrino scatterings
$\nu_\tau\nu_\tau\to f f$ and $\nu_\tau\to\nu_\tau f$,
the latter process being allowed due to background matter effects. The
familon luminosities due to these processes are given in
Ref.~\cite{PRD42-293}:

\begin{eqnarray}
Q_f(\nu_\tau\nu_\tau\to f f) &\approx&
   8.8\times 10^{63} \text{ erg/sec} \times 
   \left(\frac{m_{\nu_\tau}}{\mbox{MeV}}\right)^2
   \left(\frac{\text{GeV}}{F^L_{\nu_\tau\nu_\tau}}\right)^4 \ ,
 \label{Q_f1} \\
Q_f(\nu_{\tau}\to\nu_{\tau} f) &\approx&
 \left\{ \begin{array}{l}
   \displaystyle 1.6\times 10^{54} \text{ erg/sec} \times 
   \left(\frac{\text{MeV}}{m_{\nu_\tau}}\right)^4
    \left(\frac{\text{GeV}}{F^L_{\nu_\tau\nu_\tau}}\right)^2 \; , \
     m_{\nu_\tau}\geq 95 \text{ keV}   \\
   \displaystyle 1.9\times 10^{60} \text{ erg/sec} \times 
 \left(\frac{m_{\nu_\tau}}{\mbox{MeV}}\right)^2 
 \left(\frac{\mbox{GeV}}{F^L_{\nu_\tau\nu_\tau}}\right)^2 \; , \
     m_{\nu_\tau}\leq 95 \text{ keV} \ . \end{array} \right.
\label{Q_f3}
\end{eqnarray}
If {\em either one} of the above luminosities is larger than $\approx
10^{53}$~erg/sec, the cooling process of the supernova may be
dominated by familon emission, and the duration time of the neutrino
pulse becomes shorter than $O(1\text{ sec})$. Imposing the constraint
that the familon luminosities given in Eqs.~(\ref{Q_f1}) and
(\ref{Q_f3}) are smaller than $3\times 10^{53}$~erg/sec, we obtain the
following constraints:
\begin{eqnarray}
\nu_\tau\nu_\tau\to f f &:\;\;& \hspace{1.1em} 
\left(\frac{m_{\nu_\tau}}{\text{MeV}}\right) 
\left(\frac{\text{GeV}}{F^L_{\nu_\tau\nu_\tau}}\right)^2
     \leq 5.8\times 10^{-6} \ ,\label{constraint_1}  \\
\nu_\tau\to\nu_\tau f &:\;\;&  \left\{ 
 \begin{array}{rcll} \displaystyle
 \left(\frac{\text{MeV}}{m_{\nu_\tau}}\right)^2 
 \left(\frac{\text{GeV}}{F^L_{\nu_\tau\nu_\tau}}\right) 
 & \leq & 0.43 \;, & m_{\nu_\tau}\geq 95 \text{ keV} \\[2ex]
\displaystyle
\left(\frac{m_{\nu_\tau}}{\text{MeV}}\right) 
\left(\frac{\text{GeV}}{F^L_{\nu_\tau\nu_\tau}}\right)
 & \leq & 4.0\times 10^{-4} \;, & m_{\nu_\tau}\leq 95 \text{ keV} \ .
 \end{array} \right. \label{constraint_3}
\end{eqnarray}
Familon volume emission is sufficiently small when both
Eqs.~(\ref{constraint_1}) and (\ref{constraint_3}) are satisfied.

On the other hand, if familon interactions are strong enough, familons
effectively scatter off the neutrinos in the background and get
thermalized and trapped in the supernova. Once this happens, a thermal
sphere of familons is formed, just like the thermal neutrino sphere,
and familons can only be emitted from the surface. The familon
luminosity essentially obeys the formula of blackbody emission with
the surface temperature of the familon sphere.  The important point is
that, once the familon is trapped, the familon luminosity {\em
decreases} as the familon interaction becomes stronger.  This can be
understood in the following way: as the familon interaction becomes
stronger, familons can be thermalized with a lower temperature.
(Notice that the scattering rate increases for higher temperature.)
The surface temperature of the familon sphere then decreases, and
hence the luminosity is suppressed. Therefore, the familon luminosity
can be small enough when the scale $F$ is sufficiently low.  Following
Ref.~\cite{PRD42-293} we find that the cooling through familon
emission is sufficiently suppressed ({\em i.e.}, is less than $3
\times 10^{53}\text{ erg/sec}$) when {\em either one} of the following
constraints is satisfied:

\begin{eqnarray}
\begin{array}{l}
 f\nu_{\tau} \to  f \nu_{\tau} \\
 f f\to \nu_{\tau}\nu_{\tau} \end{array} &:\;\;& 
\displaystyle
\left(\frac{m_{\nu_\tau}}{\text{MeV}}\right) 
\left(\frac{\text{GeV}}{F^L_{\nu_\tau\nu_\tau}}\right)^2 
  \geq 8.3\times 10^{-4} \ ,
 \label{lowerbound_1}
 \\
 f \nu_{\tau} \to \nu_{\tau}  &:\;\;& \left\{ \begin{array}{l}
\displaystyle
\left(\frac{m_{\nu_\tau}}{\text{keV}}\right) 
\left(\frac{\text{GeV}}{F^L_{\nu_\tau\nu_\tau}}\right)
     \geq 85\; , \ m_{\nu_\tau}\leq 1 \text{ keV} \\
\displaystyle
\left(\frac{m_{\nu_\tau}}{\text{keV}}\right)^{1/2} 
\left(\frac{\text{GeV}}{F^L_{\nu_\tau\nu_\tau}}\right) 
     e^{-m_{\nu_\tau}/2.4\text{ keV}}
     \geq 50 \; , \ m_{\nu_\tau}\geq 1 \text{ keV} \ . 
\end{array} \right.
\label{lowerbound_3} \end{eqnarray} 
Of course, if the familon has strong interactions with other light
particles (the photon, electron, or neutron), familons may be trapped
by other processes as well. This gives additional regimes where the
earlier constraints of Eqs.~(\ref{constraint_1}) and
(\ref{constraint_3}) can be evaded.

\begin{figure}[t]
 \centerline{\epsfxsize=0.6\textwidth
 \epsfbox{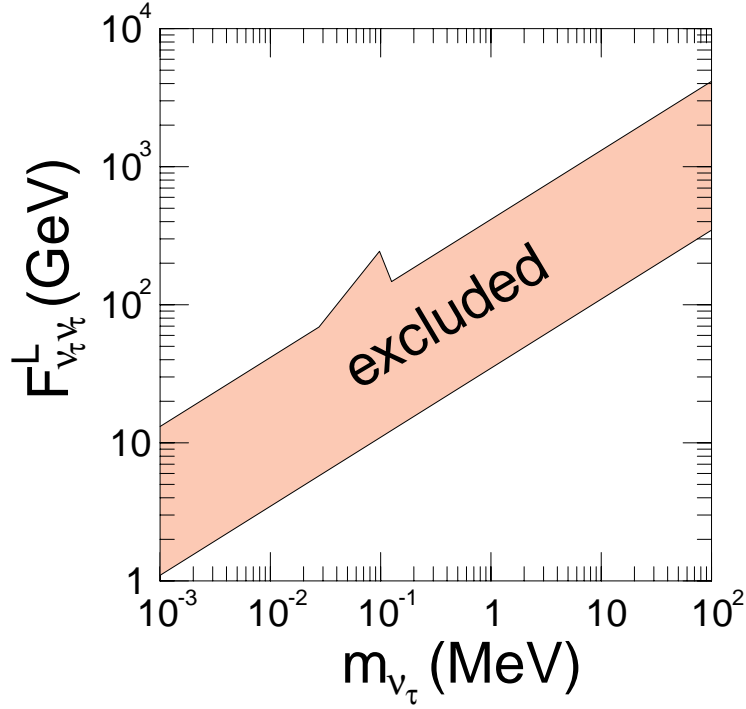}} \caption{Excluded
 region in $F^L_{\nu_\tau\nu_\tau}$ as a function of the neutrino mass
 $m_{\nu_\tau}$, derived from SN1987A.  The bounds shown correspond to
 the case of diagonal familon coupling.  For the off-diagonal case the
 bounds on $F^L_{\nu_{\tau} \nu_{\mu}}$ are very similar: the excluded
 region is only marginally shifted down by a factor of 1.2, and the
 little bump from $\nu_\tau \to \nu_\tau  f$ is absent.}
 \label{fig:sn}
\end{figure}

In Fig.~\ref{fig:sn}, we show the upper and lower bounds on the
diagonal coupling $F^L_{\nu_{\tau}\nu_{\tau}}$ as a function of the
neutrino mass $m_{\nu_{\tau}}$.  The region above the upper line is
allowed because familon emission is sufficiently suppressed by the
flavor scale $F$.  This line is basically determined by
Eq.~(\ref{constraint_1}); the slight bump is due to
Eq.~(\ref{constraint_3}).  As we can see, the lower bound on $F$ is at
most 1 TeV for the maximum allowed value of the tau neutrino mass
(18.2~MeV), and it becomes less stringent as the mass becomes
smaller. The lower boundary is determined by Eq.~(\ref{lowerbound_1}),
which supercedes Eq.~(\ref{lowerbound_3}).  The region below this line
is also allowed because the familon is trapped in the core and the
contribution to the cooling is again sufficiently small.

\subsubsection*{Familon with off-diagonal coupling}
Here we discuss SN 1987A constraints on a familon which has only an
off-diagonal coupling, such as in the case of an O(2) family
symmetry. The low-energy Lagrangian of the model can be written as

\begin{equation}
 {\cal L}_{ f} = \frac{i}{F} g_L^{\nu_\tau \nu_\mu}
 \partial_\mu f\,
 (\bar{\nu}_\tau\gamma^\mu P_{L}\nu_\mu - \bar{\nu}_\mu\gamma^\mu
 P_{L}\nu_\tau) \ .
\label{L_off}
\end{equation}
The inclusion of $\nu_e$ in the discussion is straightforward.  The
familons are produced by $\nu_\tau\nu_\tau\to f f$ via
$t$-channel $\nu_\mu$ exchange, $\nu_\mu\nu_\mu\to f f$ via
$t$-channel $\nu_\tau$ exchange, or the decays
$\nu_\tau\to\nu_\mu f$.  Following Ref.~\cite{PRD42-293} again,
\begin{eqnarray}
Q_f(\nu_\tau\nu_\tau\to f f) &\approx&
   2.2\times 10^{63} \text{ erg/sec} \times 
   \left(\frac{m_{\nu_\tau}}{\mbox{MeV}}\right)^2 
   \left(\frac{\mbox{GeV}}{F^L_{\nu_\tau\nu_\mu}}\right)^4 \ ,
 \label{Q_f4} \\
Q_f(\nu_\mu\nu_\mu\to f f) &\approx& 
   2.2\times 10^{63} \text{ erg/sec} \times 
   \left(\frac{m_{\nu_\tau}}{\mbox{MeV}}\right)^2 
   \left(\frac{\mbox{GeV}}{F^L_{\nu_\tau\nu_\mu}}\right)^4 \ ,
 \label{Q_f5} \\
Q_f(\nu_\tau\to\nu_\mu  f) &\approx&
   3.0\times 10^{62} \text{ erg/sec} \times 
   \left(\frac{m_{\nu_\tau}}{\text{MeV}}\right)^4 
   \left(\frac{\text{GeV}}{F^L_{\nu_\tau\nu_\mu}}\right)^2 \ . 
\label{Q_f6}
\end{eqnarray}
We require that {\em all} of these familon luminosities are smaller
than $3\times 10^{53}$~erg/sec, and obtain the following constraints:

\begin{eqnarray}
\nu_\tau\nu_\tau\to f f \ ,\ \nu_\mu\nu_\mu\to
 f f &:\;\;& 
\displaystyle
\left(\frac{m_{\nu_\tau}}{\text{MeV}}\right)
\left(\frac{\text{GeV}}{F^L_{\nu_\tau\nu_\mu}}\right)^2
     \leq 8.3\times 10^{-6} \ ,\label{constraint_4}  \label{cool1} \\
\nu_\tau\to\nu_\mu  f &:\;\;&
\displaystyle
 \left(\frac{\tau_{\nu_\tau}}{\text{sec}}\right) 
 \left(\frac{\text{MeV}}{m_{\nu_\tau}}\right) 
     \geq 3.3\times 10^{-5} \ , \label{constraint_6}
\end{eqnarray}
where the lifetime of $\nu_\tau$ is given by 
\begin{eqnarray}
\tau_{\nu_\tau}^{-1} = 
\frac{1}{16\pi} \frac{m_{\nu_\tau}^3}{F_{\nu_\tau \nu_\mu}^{L\; 2}} \ .
\label{tau_nu}
\end{eqnarray}

Increasing the interaction strength further into the excluded region,
familons eventually become trapped and rendered harmless again.  This
occurs when {\em any one} of the following constraints are satisfied:
\begin{eqnarray}
\begin{array}{ll}
 f\nu_{\mu} \to  f \nu_{\mu}\, , & 
 f\nu_{\tau} \to  f \nu_{\tau}\, , \\
 f f\to \nu_{\mu}\nu_{\mu}\, , & 
 f f\to \nu_{\tau}\nu_{\tau} \end{array} &:\;\;& 
\displaystyle
\left(\frac{m_{\nu_\tau}}{\text{MeV}}\right) 
\left(\frac{\text{GeV}}{F^L_{\nu_\tau\nu_\mu}}\right)^2 
  \geq 1.2\times 10^{-3} \ ,
 \label{lowerbound_4}
 \\
 f\nu_{\mu} \to \nu_{\tau}  &:\;\;& (\tau_{\nu_\tau}/\text{sec}) 
\displaystyle
     \left(\frac{\text{MeV}}{m_{\nu_\tau}}\right) \leq 10^{-6} \ .
\label{lowerbound_5}
\end{eqnarray}
As mentioned before, if the familon has strong interactions with other
light particles, these interactions may lead to thermalization of
familons as well, resulting in additional allowed regions for low
flavor scales.

The resulting excluded region is fairly similar to that of the
diagonal coupling case.  The constraint from the decay process
Eq.~(\ref{constraint_6}) is important only for smaller
$F^L_{\nu_\tau\nu_\mu}$ or larger masses $m_{\nu_\tau}$, values that
are outside our range of interest.  In addition, the small bump in
Fig.~\ref{fig:sn} now disappears due to the absence of the $\nu_\tau
\to \nu_\tau  f$ process.  The dominant constraints are therefore 
from Eqs.~(\ref{cool1}) and (\ref{lowerbound_4}) in the off-diagonal
case, which differ from the dominant constraints of
Eqs.~(\ref{constraint_1}) and (\ref{lowerbound_1}) in the diagonal
coupling case only by a small constant factor.  The boundary of the
excluded region is therefore given by the lines of Fig.~\ref{fig:sn}
shifted downwards by a factor of 1.2 in $F$.

\subsection{Bounds from loop-induced couplings}
\label{sec:loop}

In the previous subsection, we considered astrophysical constraints on
tree-level familon couplings to $\nu_{\tau}$.  In addition, however,
astrophysical bounds may also be used to constrain familon couplings
to all other particles, as these couplings may induce couplings of
familons to electrons and nucleons at the loop level.  While these
induced couplings are suppressed by the usual loop factors, the bounds
on familon couplings to first generation particles are so stringent
that these constraints may be strong in certain cases.  In fact, we
will see below that the contributions to induced couplings are
proportional to fermion masses, and so these constraints are
particularly relevant for couplings of familons to third generation
fermions. In this subsection, we will estimate the induced couplings
for various choices of the family symmetry group and determine what
lower bounds on flavor scales $F$ result from current astrophysical
constraints.  For simplicity, we will limit our discussion here to
familons with flavor-diagonal couplings to the third generation, and
ignore possible rotations relating the flavor and mass eigenstates.
Extensions of this analysis to more general cases are straightforward.

\begin{figure}
\centerline{\epsfxsize=0.3\textwidth \epsfbox{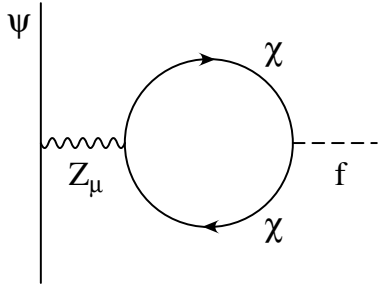}}
\caption{The $Z$-$ f$ mixing diagram.}
\label{fig:Feynman}
\end{figure}

To evaluate the strength of the induced coupling, we will begin by
considering the low energy effective theory below the flavor scale
$F$.  In this approach, the theory is specified by the flavor
symmetry, that is, the low energy derivative couplings of the familon,
and no further knowledge of the mechanisms of flavor symmetry breaking
is required.  With the assumptions given above, the dominant
contribution to the induced couplings is from the $Z$-$ f$ mixing
graph shown in Fig.~\ref{fig:Feynman}.  Here $\chi$ is any one of the
third generation particles directly coupled to the familon, and $\psi
= e$, $u$, or $d$.  (There are also additional contributions from
penguin-like $W$ diagrams, but these are suppressed by mixing angles,
{\em e.g.}, $V_{td}^2$ in the case of $\psi=d$.)  Let us define the
$ f$-$\chi$ coupling as

\begin{equation}
\frac{1}{F} \partial_{\mu}  f \, \bar{\chi} \gamma^{\mu} (g_L P_L + g_R
P_R) \chi \ ,
\end{equation}
and the $Z$-$\chi$ coupling as

\begin{equation}
Z_{\mu} \bar{\chi} \gamma^{\mu} ( g_L^Z P_L + g_R^Z P_R ) \chi \ ,
\end{equation}
where $g_L^Z = g_Z (I_\chi - Q_\chi \sin^2 \theta_W)$, $g_R^Z = -g_Z
Q_\chi \sin^2 \theta_W$, and $g_Z = e/(\sin\theta_W \cos\theta_W)$.
The induced $Z$-$ f$ mixing from the fermion loop is divergent, and
the logarithmically-enhanced contribution is

\begin{eqnarray}
{\cal L}_{Z- f} &=& \frac{2N_c}{(4\pi)^2 F} \ln
\frac{\Lambda^2}{m_\chi^2} \left[ m_\chi^2 (g_L - g_R ) (g_L^Z - g_R^Z)
g^{\mu\nu} - \frac{1}{3} (g_L g_L^Z + g_R g_R^Z) (p^2 g^{\mu\nu} -
p^{\mu} p^{\nu} ) \right] Z_{\mu} \partial_{\nu}  f \nonumber\\
&=& \frac{2N_c}{(4\pi)^2 F} m_\chi^2 \ln \frac{\Lambda^2}{m_\chi^2} 
(g_L - g_R ) (g_L^Z - g_R^Z) Z^{\mu} \partial_{\mu}  f \ ,
\label{zphimix}
\end{eqnarray}
where $N_c$ is the number of colors of the fermion $\chi$, and
$\Lambda$ is the effective ultraviolet cutoff of the order of the
flavor scale $F$.  Note that $g_L^Z - g_R^Z = g_Z I_\chi$ and the
fermion charge $Q_\chi$ drops out: the Ward--Takahashi identity
guarantees that the $Q_\chi \sin^{2} \theta_{W}$ piece in the $Z$
vertex gives a completely transverse vacuum polarization amplitude
proportional to $p^{2} g^{\mu\nu} - p^{\mu} p^{\nu}$, which vanishes
when contracted with $\partial_{\mu} f$.  The leading contributions to
the induced couplings are determined by the amount of current
non-conservation, {\em i.e.}, the masses of the particles in the loop,
and the third generation couplings therefore give the most important
contributions.\footnote{Note that the radiatively-induced mixing
operator can be written in the manifestly gauge invariant form
$i(H^{\dagger} D_{\mu} H - D_{\mu} H^{\dagger} H) \partial^{\mu}
f$. This operator may be present at tree-level if the familon couples
to ``Higgs number,'' but we will not consider this case.}

The mixing of Eq.~(\ref{zphimix}) is logarithmically-enhanced and so
typically gives the leading contribution if present.  However, there
are cases in which this term is not present or is highly suppressed.
First, it may be that the amount of current non-conservation is itself
suppressed by inverse powers of the flavor scale. For instance, in the
singlet Majoron model~\cite{CMP}, lepton number conservation in the
low-energy theory is violated only by neutrino masses which are of
order $1/F$ due to the seesaw mechanism.  The neutrino loop
contribution to the $Z$-Majoron mixing is then $1/F^3$ and highly
suppressed.  Second, if the familon coupling is vector-like so that
$g_L = g_R$, the logarithmically-enhanced term is absent.  For
example, a familon coupled to the (possibly generation-dependent)
baryon number current has this property. Finally, this mixing is also
absent if the familon has only flavor off-diagonal couplings.  In all
of these cases, the contribution of Eq.~(\ref{zphimix}) is absent or
suppressed, and the leading contributions to $Z$-$ f$ mixing come
from non-logarithmically-enhanced threshold corrections which may be
of order $1/F$.  Such corrections are sensitive to physics at the
flavor symmetry breaking scale, and are therefore model-dependent.

With these caveats in mind, we now assume that the leading
contribution given in Eq.~(\ref{zphimix}) is present, and determine
bounds on $F$ for various flavor symmetries.  The
logarithmically-enhanced mixing induces an effective coupling

\begin{eqnarray}
{\cal L}_{\text{eff}} &=& \frac{2 N_c}{(4\pi)^{2}F} g_Z^2 I_\chi 
(g_L-g_R) \frac{m_\chi^2}{m_Z^2} \ln \frac{\Lambda^{2}}{m_\chi^2} 
      \partial_{\mu}  f \, \bar{\psi} \gamma^{\mu}
      \left( I_{\psi}P_L - Q_{\psi} \sin^2 
        \theta_{W}\right) \psi \nonumber \\
&=& i \frac{2 N_c}{(4\pi)^{2}F} g_Z^2 I_\chi (g_L - g_R)
\frac{m_\chi^2}{m_Z^2} \ln \frac{\Lambda^{2}}{m_{\chi}^{2}}  
       I_{\psi} m_{\psi} f\bar{\psi} \gamma_{5} \psi \ ,
\label{bracketed}
\end{eqnarray}
where in the last step we have integrated by parts and substituted the
equations of motion for $\psi$.  For example, for a familon coupled to
$t_R$, $N_c = 3$, $I_\chi = 1/2$, and $g_L - g_R = -1$.  For a familon
coupled to $Q_L$, the contributions from both $t_L$ and $b_L$ must be
summed.

The effective coupling of Eq.~(\ref{bracketed}) is constrained from
various sources.  For the case $\psi = e$, a stringent constraint is
provided by red giants.  Familon-electron couplings lead to additional
sources of red giant cooling, which, if too large, would destroy the
agreement between the observed population of red giants in globular
clusters and stellar evolution theory.  Such constraints have been
studied extensively in the literature~\cite{redgiants,RW}.  The
current best upper limit on the coupling is~\cite{RW}

\begin{equation}
        g < 2.5 \times 10^{-13} 
\end{equation}
for ${\cal L}_{\text{eff}} = -i g  f \bar{e} \gamma_{5}
e$.\footnote{For a larger coupling, the familon may be trapped in red
giants and not contribute to their cooling.  Still, they can be
emitted from the Sun and change its dynamics significantly.  For yet
larger couplings, familons may be trapped in the Sun as well, but then
they contribute to the thermal transport.  Combination of these
constraints exclude all couplings to electrons larger than this one.
See Ref.~\cite{PRep198-1} for further details.}  The strongest bound
on the family symmetry breaking scale is for familon couplings that
are dominantly proportional to $m_t^2$, as, for example, when a
familon is coupled only to $t_R$.  Such a case results in the bound
\begin{equation}
        F^R_{t t} > 1.2 \times 10^{9}\text{ GeV} \ ,
\end{equation}
where we have taken $\Lambda = F$. Weaker, but still significant
constraints are obtained if the familon coupling is dominated by
$m_b$, as when the familon couples only to $b_R$. The bound in
this case is

\begin{equation}
F^R_{bb} > 6.1 \times 10^{5}\text{ GeV} \ .
\label{frbb}
\end{equation}
Notice that, in the case where the familon couples to $t_R$ and $Q_L$
with the same charge, the bound of Eq.~(\ref{frbb}) also holds for the
corresponding flavor scales. However, possibly stronger bounds may
also be possible if model-dependent non-logarithmically enhanced terms
proprtional to $m_t^2$ are present. If the familon contribution to the
induced coupling is dominantly proportional to $m_{\tau}^{2}$, we find
the constraint
\begin{equation}
F^R_{\tau\tau} > 2.5 \times 10^{4}\text{ GeV} \ .
\end{equation}

Similar bounds may be obtained from induced familon couplings to
nucleons using constraints from supernova SN~1987A by rescaling the
bounds on axion couplings.  These constraints are somewhat more
ambiguous because of the loss of coherence in axion emission due to
nucleon spin fluctuations caused by scattering effects in the
supernova core~\cite{RS}.  More realistic estimates were addressed in
Refs.~\cite{realistic,PRD56-2419}.  The constraints yield results
comparable to the red giant bound on electron couplings, but with
larger error bars.

Finally, we stress again that these bounds are for specific flavor
symmetries.  For certain examples mentioned above, the
logarithmically-enhanced contribution to the induced coupling is
absent.  For such flavor symmetries, the threshold corrections at the
flavor scale must be studied separately for each model, and the flavor
scale can be constrained only after the model-dependent coefficients
are known.  However, from the numerical estimates above, it is clear
that the induced loop-level bounds can provide interesting constraints
on flavor-diagonal familon couplings. Such bounds are particularly
interesting for couplings to the third generation, as they are
enhanced for large fermion masses. Note also that couplings to the top
quark are stringently bounded and are extremely difficult to bound by
other means.

\subsection{Bounds from effects induced by flavor-mixing}
\label{sec:mixing}

In this section, we have so far parametrized and constrained possible
familon couplings individually by introducing effective flavor scales,
neglecting possible mismatches between flavor and mass eigenstates.
However, as noted in Sec.~\ref{sec:familons}, when bounds on a
particular familon coupling are very stringent, such as in the case of
bounds on familon couplings to the first generation from supernovae,
one can also obtain interesting bounds on other familon couplings from
flavor-mixing effects.  In this subsection, we will consider such
bounds in the quark sector. In extensions of the standard model with
massive neutrinos, similar arguments hold in the leptonic
sector. Additional constraints may also be obtained if the gauge
symmetry is enlarged.

Let us assume that the flavor and mass eigenstates coincide in the
down sector. A generic familon coupling term for left-handed down-type
quarks is then
 
\begin{equation}
 {\cal L}_{ f} =
 \frac{1}{F^L_{IJ}} \partial_\mu  f \, 
\bar{d}_I \gamma^\mu P_L d_J \ ,
\end{equation}
where $I$ and $J$ are generational indices. Notice that, in this
section, the familon $ f$ is a real scalar for the diagonal
couplings ($I=J$), and a complex scalar for the off-diagonal ones
($I\neq J$).  (Thus, in the off-diagonal case, there is also a
Hermitian conjugate term in the Lagrangian.)

The up-type coupling required by gauge invariance in terms of down
quark mass eigenstates is
 \begin{eqnarray}
 {\cal L}_{ f} &=&
 \frac{1}{F_{IJ}^L} \partial_\mu  f\, V^*_{iI}\bar{u}_i
 \gamma^\mu P_L V_{jJ} u_j 
 \nonumber \\ &\equiv& 
 \frac{1}{2F_{IJ}^L}
 (x_u\partial_\mu  f\, \bar{u}\gamma_\mu\gamma_5 u + \cdots) \ ,
 \label{phi-q-q}
 \end{eqnarray}
 where $x_u=-V^*_{uI}V_{uJ}$. Therefore, a constraint on $F^L_{uu}$
obtained from supernova cooling through familon--nucleon coupling
implies similar constraints on the expressions
$F_{bs}^L/(V^*_{ub}V_{us})$, $F_{bd}^L/(V^*_{ub}V_{ud})$, and so on.
Of course, different contributions to the same effective coupling
${F^L_{uu}}^{-1}$ may also have opposite signs, which must be checked
in the specific model under investigation.  

To derive constraints on the flavor symmetry breaking scales, we must
convert the quark level couplings to the effective nucleon-familon
couplings of the form ${\cal L}_{\text{int}}\sim ig_{ f
NN} f\bar{N}\gamma_5 N$. This can be done through a generalized
Goldberger-Treiman relation. With the interaction given in
Eq.~(\ref{phi-q-q}), we obtain

\begin{eqnarray}
 g_{ f NN} = \frac{m_N}{F_{IJ}^L} x_u \Delta^{(N)}_u \ ,
 \label{phi-N-N}
\end{eqnarray}
where $m_N\simeq 0.94$ GeV is the nucleon mass, and the coefficients
$\Delta^{(N)}_q$ are given by \cite{PRD56-2419}

\begin{eqnarray}
 \Delta^{(p)}_u &\simeq& \Delta^{(n)}_d \simeq 0.80 \ ,
 \nonumber \\ 
 \Delta^{(n)}_u &\simeq& \Delta^{(p)}_d \simeq -0.46 \ ,
 \nonumber \\ 
 \Delta^{(p)}_s &\simeq& \Delta^{(n)}_s \simeq -0.12 \ .
 \nonumber
\end{eqnarray}
Here, we have assumed that the flavor symmetry is anomaly-free for
SU(3)$_C$. If the flavor symmetry is anomalous under SU(3)$_C$, there
are anomaly-induced contributions to Eq.~(\ref{phi-N-N}); see
Refs.~\cite{PRep198-1,PRD56-2419,aph9707268} for discussions of
constraints on axions.
 
The effective couplings of Eq.~(\ref{phi-N-N}) are constrained by
supernova SN 1987A.  In Ref.~\cite{PRD56-2419}, the upper bound on
$g_{ f pp}$ is given as a function of $g_{ f nn}/g_{ f
pp}$.  For simplicity, we adopt the most conservative constraint on
$g_{ f pp}$,

\begin{eqnarray} g_{ f
pp}\lesssim 3\times 10^{-10} \ , 
\end{eqnarray} 
and use this to estimate bounds on the flavor symmetry breaking
scales.

Under the assumption that only one down-type familon coupling exists
at a time, that is, ignoring possible cancellations between two
different contributions, we find bounds on third-generation couplings

\begin{eqnarray}
 F^L_{bb} & > & 3 \times 10^4 \text{ GeV}
 \left(\frac{\left|V_{ub}\right|}{3.5\times 10^{-3}}\right)^2
 \left(\frac{\Delta^{(p)}_u}{0.80}\right) \ ,
 \\
 F^L_{bs} & > & 3 \times 10^6 \text{ GeV}
 \left(\frac{\left|V_{ub}\right|}{3.5\times 10^{-3}}\right)
 \left(\frac{\left|V_{us}\right|}{0.22}\right)
 \left(\frac{\Delta^{(p)}_u}{0.80}\right) \ ,
 \\
 F^L_{bd} & > & 1 \times 10^7 \text{ GeV}
 \left(\frac{\left|V_{ub}\right|}{3.5\times 10^{-3}}\right)
 \left(\frac{\left|V_{ud}\right|}{0.98}\right)
 \left(\frac{\Delta^{(p)}_u}{0.80}\right) \ .
\end{eqnarray}
 An even stronger constraint is obtained for $F^L_{sd}$:

\begin{eqnarray}
 F^L_{sd} >  8 \times 10^8 \text{ GeV}
 \left(\frac{\left|V_{us}\right|}{0.22}\right)
 \left(\frac{\left|V_{ud}\right|}{0.98}\right)
 \left(\frac{\Delta^{(p)}_u}{0.80}\right) \ .
\end{eqnarray}
This constraint is, however, weaker than the laboratory bound.

Note that the above bounds are obtained under the assumption that the
mass and flavor eigenstates are identical in the down sector.  If, on
the other hand, these eigenstates were assumed to be identical in the
up sector, familon coupling to the $s$- and $d$-quark arises due to
the mixing effect, and the supernova constraints on $F^L_{dd}$ and
$F^L_{ss}$ could be used instead. In particular, an interesting bound
is derived for $F^L_{tt}$:

\begin{eqnarray}
 F^L_{tt} >  7 \times 10^5 \text{ GeV}
 \left(\frac {\left|V^*_{ts}V_{ts}\Delta^{(p)}_s 
            +       V^*_{td}V_{td}\Delta^{(p)}_d\right|} 
             { (4.1\times 10^{-2})^2\times 0.12   
            +  (5.7\times 10^{-3})^2\times 0.46 }\right)\ .
\end{eqnarray}
This bound is about one order of magnitude stronger than the bound on
$F^L_{bb}$ since, in this case, the effective familon coupling to the
nucleon is dominated by the $s$-quark contribution, and the
interaction is therefore not as highly Cabbibo-suppressed as in the
$F^L_{bb}$ case.

Of course, there is no reason why the familon coupling is diagonalized
in one sector.  However, as the up and down sectors cannot
be diagonalized in the same basis, one generally expects similar
mixing-induced constraints in all cases.

\section{Prospects for Future Probes: $\protect\bbox{B}$ Factories}
\label{sec:Bfactories}

In this section, we estimate what constraints on familon couplings may
be obtained from the current CLEO data set and the upcoming CLEO III,
BABAR and BELLE experiments.  We make no attempt to conduct detailed
experimental studies appropriate to each of these experimental
settings.  Rather, our intent here is to describe a number of analyses
that are likely to significantly improve the present limits on familon
energy scales, and, we believe, merit further study.  We will begin
with investigations of hadronic couplings of $b$ quarks to familons,
and then consider decays of the $\tau$ lepton to familons.

\subsection{Bounds from $\protect\bbox{B}$ decays}
\label{sec:bbounds}

In all of the experiments mentioned above, one may search for the
exclusive decays $B^{\pm} \to (\pi^{\pm}, K^{\pm})  f$ and $B^0
\to K_s  f$.  These exclusive modes have smaller branching
fractions than the inclusive modes $b\to (d,s) f$, but have clear
experimental signatures due to their simple two-body kinematics.  The
form factor of
\begin{equation}
  \langle K^-(p')| \bar{s}\gamma^{\mu} b(q^{2} = 0) |B^-(p)\rangle
  = F_{1}(0) (p+p')^{\mu} \ ,
\end{equation}
which is necessary to calculate branching fractions, has been
estimated by Colangelo {\em et al.}~\cite{Colangelo} to be $F_1(0) =
0.25\pm 0.03$, based on sum rules. Estimates of $F_1(0)$ based on the
quark model are 0.34, 0.36, 0.30, or 0.35, depending on which quark
model parameters are assumed~\cite{quark}.  We could not find other
estimates of this particular form factor in the literature, but
various estimates for $B \to \pi$ transitions give comparable but
slightly larger values.  This is reassuring, since they must agree in
the flavor symmetric limit.  The decay rate $B\to K f$ is given by

\begin{equation}
  \Gamma(B\to K f) = \frac{1}{16\pi}\frac{m_B^3}{F^2}g_V^2 \beta^3
  |F_1 (0)|^2  \ ,
\end{equation}
where $\beta = 1-m_K^2/m_B^2$.  (If the coupling is purely axial,
there is, of course, no contribution to $B\to K f$; searches for
decays to $K^{*} f$ are required to bound such couplings.)
Neglecting the mass difference between the $b$ quark and $B$ meson,
and using the naive spectator model for the $B$ meson decay, one finds

\begin{equation}
  \frac{\Gamma(B\to K f)}{\Gamma(b\to s f)}
  \approx |F_{1}(0)|^{2} \frac{g_{V}^{2}}{g_{V}^{2} + g_{A}^{2}}\ .
\end{equation}

The concept of the search for such exclusive decay modes is relatively
simple.  After applying the standard cuts to suppress continuum
$q\bar{q}$ and lepton pair events, one looks for events at the
$\Upsilon (4S)$ resonance that have either an isolated $K_s$, or an
isolated charged meson $\pi^{\pm}$, $K^{\pm}$ together with large
missing energy.  In the center-of-momentum frame, the energy of the
meson $P = \pi^{\pm}$, $K^{\pm}$, or $K_s$ must be in the narrow range
\begin{equation}
  \frac{\sqrt{s}}{4} \left[ \left( 1 + \frac{m_P^2}{m_B^2}\right)
    - \beta \left( 1 - \frac{m_P^2}{m_B^2}\right) \right] < E_P
  < \frac{\sqrt{s}}{4} \left[ \left( 1 + \frac{m_P^2}{m_B^2}\right)
    + \beta \left( 1 - \frac{m_P^2}{m_B^2}\right) \right] ,
\end{equation}
where $\beta = \sqrt{1-4m_B^2/s}=0.0645$, and $m_P$ is the mass of the
meson.  One can also require that, after excluding the isolated
energetic meson whose energy is in the above range, all the tracks and
energy deposits in the calorimeters reconstruct $m_B$ and have total
energy $\sqrt{s}/2$ in the center-of-momentum frame.

Of the existing analyses, the one most similar to that described above
is a search for $B^\pm \to l^\pm \nu_l$ by the CLEO
Collaboration~\cite{lnu}.  The reported upper bounds on the branching
fractions are $1.5\times 10^{-5}$ $(e)$, $2.1\times 10^{-5}$ $(\mu)$,
and $2.2\times 10^{-3}$ $(\tau)$.  The reach for the $P f$ mode
is expected to be worse than for $e\nu$ or $\mu\nu$, as the continuum
backgrounds are larger and the detection efficiencies are worse for
mesons.  However, we expect the sensitivity to such meson decays to be
greater than to the $\tau\nu$ mode, because the mesons have
more-or-less fixed energy, unlike in the $\tau\nu$ case.  $\pi/K$
separation is probably difficult with the current CLEO data
set~\cite{pipi}, but this analysis may still give us an upper bound on
$B(B\to \pi f)+B(B\to K f)$ somewhere at the $10^{-4}$ to
$10^{-3}$ level~\cite{Stone,Patterson}.  Such a constraint would be
competitive with the upper bound inferred from the ALEPH $b\to
s\nu\bar{\nu}$ study discussed in Sec.~\ref{sec:LEP}.

Particle identification will be much better at CLEO III, which will
allow $\pi/K$ separation, and will be even better at BABAR and
BELLE. The higher luminosity at these machines will also help, and an
upper bound of $10^{-5}$ may be possible~\cite{Stone}.  Such a bound
would imply a bound on the flavor symmetry breaking scale of
$F^V_{bd}, F^V_{bs} \agt 2 \times 10^8$~GeV!

\subsection{Bounds from $\protect\bbox{\tau}$ decays}
\label{sec:taubounds}

We now turn our attention to the lepton sector.  The decay rate for
$\tau \to l f$ is given in Eq.~(\ref{hier}).  A search for $\tau
\to l  f$ suffers from the standard model background
$\tau \to l \nu\bar{\nu}$.  A conventional method for bounding the
branching fraction to familons is to fit the momentum spectrum of the
electron (muon) from tau decay to a linear combination of the standard
model spectrum, which drops approximately linearly for large momenta, 
and a possible
contribution from the familon mode, which is flat.  The ARGUS bound
quoted in Sec.~\ref{sec:decays} was obtained by this method.  CLEO has
not reported a similar analysis.  However, in a recent CLEO analysis
of the Michel parameter in $\tau$ decays~\cite{CLEO-tau}, the electron
momentum distribution of 33531 $\tau^+ \tau^- \to (e^\pm
\nu\bar{\nu}) (\pi^\mp \pi^0 \nu)$ events was presented in uniform
0.25 GeV bins.  This distribution may be fit beautifully by the
standard model alone, and contains about 90 events in the highest
momentum bin.  For reference, the contribution of a familon decay mode
with $B(\tau \to e f) = 3\times 10^{-3}$, a branching fraction near
the current ARGUS limit, would contribute 28~events in each bin,
leading to a significant excess at high momentum.

\begin{figure}
\centerline{~\hfill
  \psfig{file=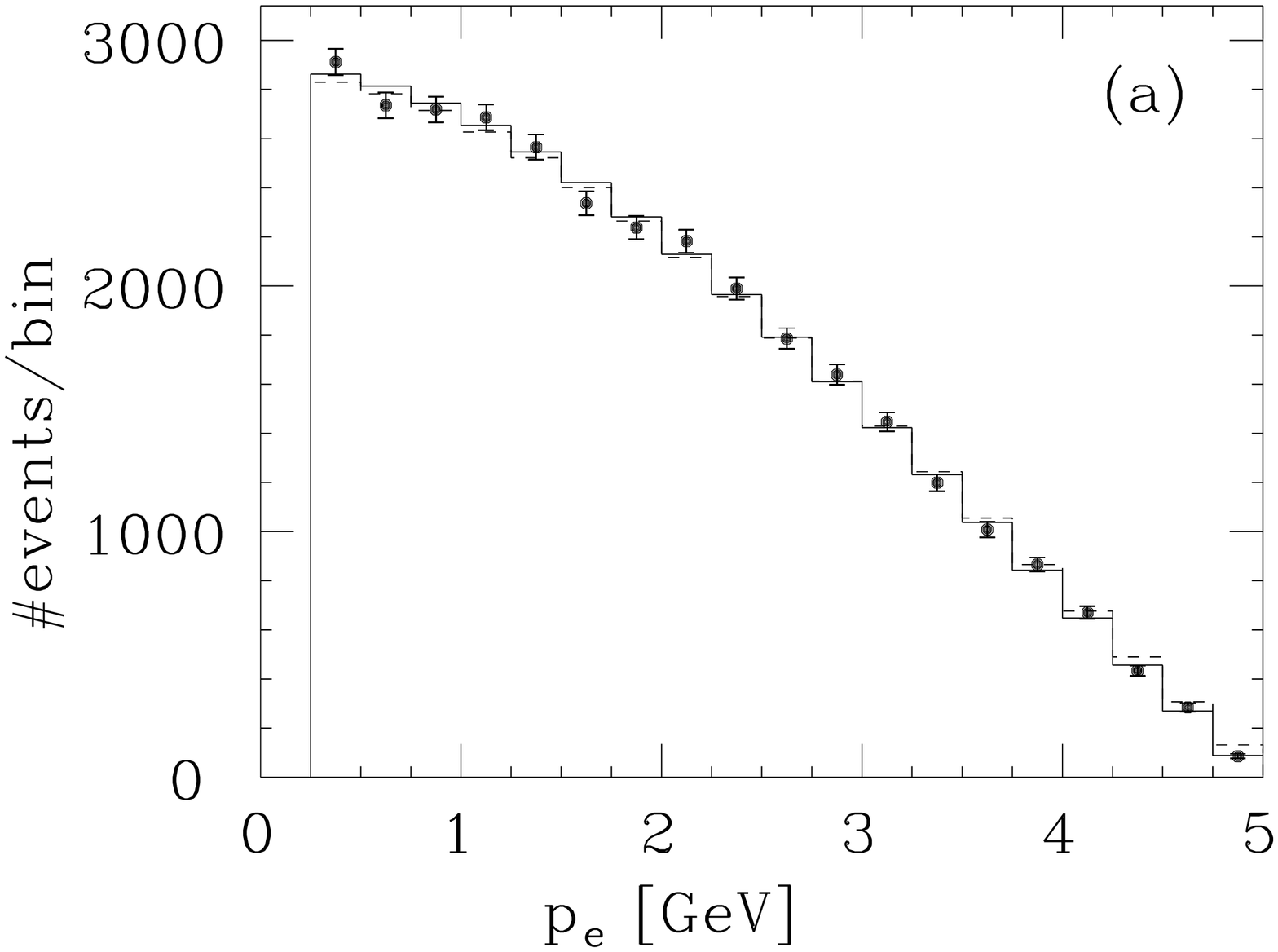,angle=90,width=0.4\textwidth}
\hfill \psfig{file=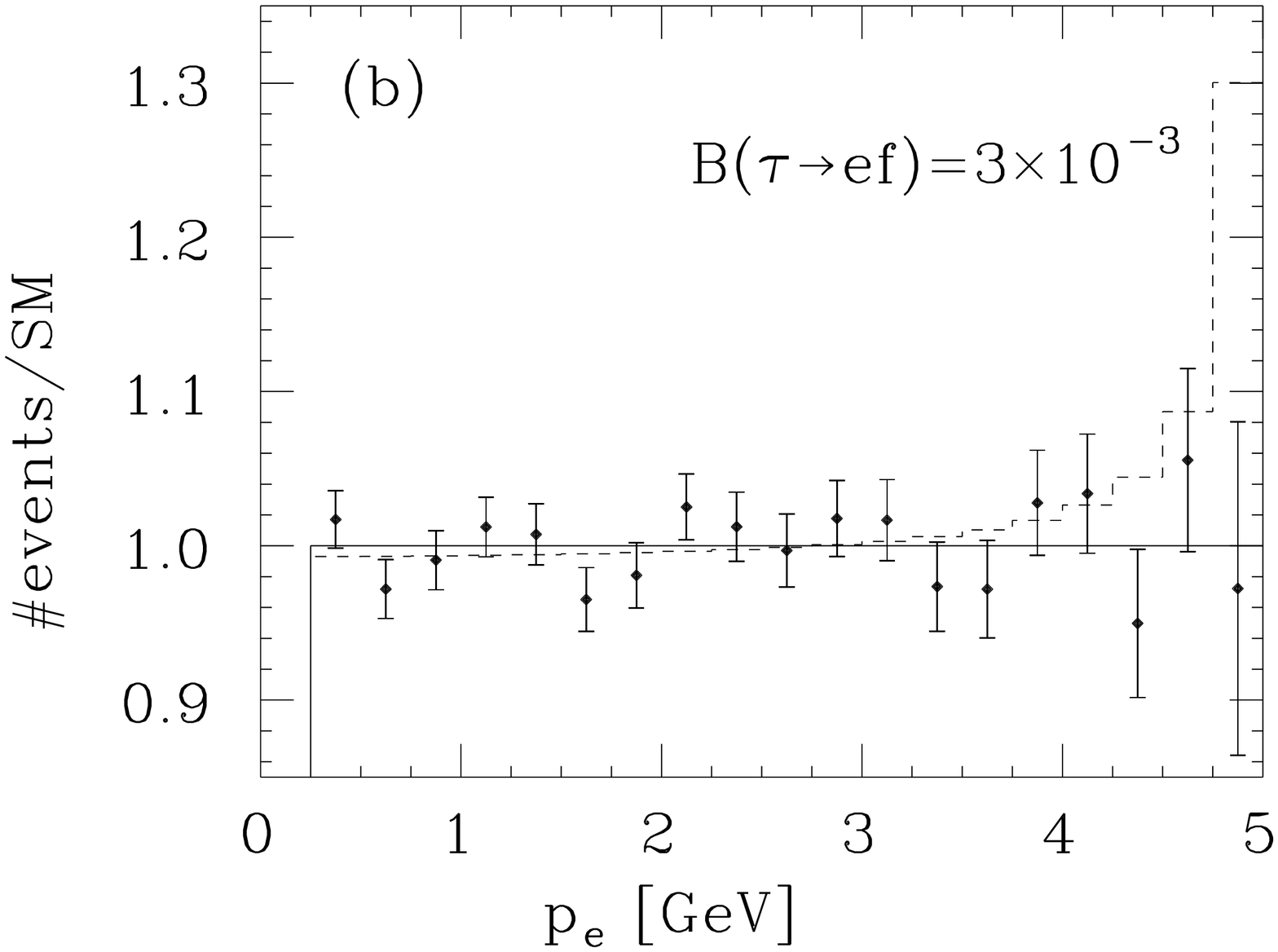,angle=90,width=0.4\textwidth}\hfill ~}
\vspace*{.2in}
\caption{(a) The $e$ momentum spectrum from $\tau$ decays.  The 
points with error bars are our Monte Carlo simulation for the standard
model only, normalized to the size of the current CLEO event
sample. The solid histogram is the standard model prediction for $\tau
\to e\nu\bar{\nu}$.  The dashed histogram is the predicted spectrum
with $B(\tau\to e f)=3\times 10^{-3}$, again normalized to the
current event sample.  (b) The same as (a) but plotted as a ratio to
the standard model prediction.}
\label{fig:emom}
\end{figure}

To see the possible sensitivity given the current CLEO data set, we
generated 33531 standard model $\tau \to e+\text{missing}$ events. The
momentum spectrum from our Monte Carlo simulation (points with error
bars), along with the predicted standard model spectrum (solid
histogram), is shown in Fig.~\ref{fig:emom}a. For comparison, we also
plot the spectrum given a hypothetical familon branching fraction of
$B(\tau \to e f) = 3\times 10^{-3}$ normalized to the same number
of events.  In Fig.~\ref{fig:emom}b, we plot the ratio to the standard
model prediction to make the familon contribution more visible.  Note
that the spectrum with the familon contribution differs considerably
in the high momentum bins.  By fitting the Monte Carlo data to the
linear combination of the standard model and familon modes, we find
that CLEO can obtain an upper bound on the familon branching fraction
of $1.6 \times 10^{-3}$ (95\% CL). This would already improve the
ARGUS bound~\cite{albrecht} on $\tau \to e f$ slightly, leading
to a lower bound of $F_{\tau e} > 7 \times 10^6$~GeV.  Note that in
this analysis, only events with the $\pi^\pm \pi^0 \nu$ decay of the
other $\tau$ were used.  This requirement was motivated in the
original study by the desire to study spin correlations between
decaying $\tau$ pairs, but is not necessary for our purpose.  The
statistical power of our analysis may therefore be boosted by
including events with additional decay modes of the other
$\tau$.\footnote{Another interesting possibility is to exploit the
spin correlations between decaying $\tau$ pairs by selecting events
with $\omega > 0$ (see Ref.~\cite{CLEO-tau} for the definition of
$\omega$).  Such a selection enhances the right-handed $\tau^-$
(left-handed $\tau^+$) decaying to leptons, and thereby suppresses the
electron momentum spectrum at the end point. The sensitivity to
familon contributions at this endpoint is then improved.}

In this analysis, the systematic effects appear to be under control.
The momentum dependence of the electron identification efficiency can
be calibrated by using the actual data, for instance, by using
radiative Bhabha events, and this calibration improves with
statistics.  In addition, the background is small in the above CLEO
data sample.  Indeed, all measurements of $\tau$ decay parameters are
statistically limited and we expect a similar situation for the
familon analysis.  Note also that in the above analysis we simply fit
to the standard model contribution, allowing its normalization to
vary.  In principle, one can determine this normalization by measuring
the efficiencies of $\tau$ identification in each decay mode through
methods analogous to the multi-tag methods employed in the measurement
of $R_b$ in $Z$ decays~\cite{ALEPHRb}.  We therefore conclude that a
dedicated analysis could well lead to an upper bound on $B(\tau \to
e f)$ below the $10^{-3}$ level.

The $\tau\to \mu f$ mode is more difficult because the muon
identification efficiency is less well-calibrated and the statistics
is slightly poorer, with 21680 events in the CLEO analysis.  Again,
however, the uncertainties are dominated by statistical
errors.\footnote{In the Michel parameter analysis, a large systematic
uncertainty arises because the standard model muon decay parameters
are not assumed.  For our purposes, however, we may assume the
standard model predictions and eliminate these uncertainties.}  We
therefore expect an upper bound on $B(\tau \to \mu f)$ only
slightly worse than that on the $e f$ mode.

Although we are concerned primarily with $B$ factories in this
section, we should note that the above analysis may also be applied at
LEP.  For example, in a recent OPAL analysis of $\tau$
polarization~\cite{OPAL}, a large sample of 25000 $\tau \to e$ events
was studied. By fitting the $e$ momentum distribution as described
above, an upper bound on $B(\tau \to l  f)$ at the $2\times
10^{-3}$ level could be derived.  (Here, we have simply scaled the
CLEO results given above to the OPAL statistics; we have checked that
the momentum distributions are sufficiently similar that such an
approximation is valid.) Combining the four LEP experiments, we expect
an upper bound of $\sim 10^{-3}$.

At the asymmetric $B$ factories, the boosted center-of-momentum system
would somewhat complicate the analysis of the electron (muon) momentum
spectrum.  We are not aware of any studies at these colliders.
However, given that the event samples available at these machines will
be much larger than the current CLEO data set, we expect BABAR and
BELLE to place constraints significantly better than the $10^{-3}$
level, and possibly at the $10^{-4}$ level.  The study of familon
decays at these machines is extremely promising, and worthy of further
study.  It must be mentioned, furthermore, that the future $B$
factories will be able to improve the upper bound on the mass of
$\nu_{\tau}$ significantly to the level of 3~MeV, or possibly even
1~MeV~\cite{Seiden}.  As we will see in the next section, the
interplay between bounds on the $\nu_{\tau}$ mass and bounds on
branching ratios to familons is very interesting from a cosmological
point of view.

Finally, we note that the bounds on $\tau\to l  f$ branching
ratios are expected to be even better at a tau-charm factory.
Ref.~\cite{Gonzalez-Garcia} has shown that one can reach the level of
$B(\tau \to e f) < 10^{-5}$ using the standard optics or even
$10^{-6}$ using a monochromator. This would raise the lower bound on
the flavor scale to $F_{\tau e} \agt 10^8$~GeV. The $\mu f$ mode
is more difficult, and is limited by the $\mu/\pi$ separation
capability~\cite{Gonzalez-Garcia}.  However, a bound better than the
$10^{-3}$ level using a RICH detector for particle identification is
expected~\cite{Gonzalez-Garcia}.
 
\section{Implications for neutrino cosmology}
\label{sec:cosmology}

Non-standard properties of neutrinos are always interesting in
cosmology, and in fact, heavy unstable neutrinos are advocated in
certain scenarios to obtain reasonable agreement between theory and
observation. The heavy neutrino is typically taken to be the tau
neutrino, and we will assume this to be the case in this section.
Once a decaying neutrino is required, its decay into a lighter
neutrino and a massless boson is the most harmless. Visible neutrino
decays are usually severely constrained from SN 1987A. As mentioned in
Sec.~\ref{sec:astrophysics}, the energy released from SN 1987A was
mostly carried away in neutrinos, and the visible luminosity of SN
1987A was much smaller. However, if neutrinos decay into visible
particles, such as photons or electrons, neutrinos emitted from SN
1987A that decay before reaching the earth may increase the apparent
visible luminosity of SN 1987A to levels much larger than
observed~\cite{prl62-509}.~\footnote{Such constraints may be evaded in
scenarios with sufficiently long-lived neutrinos~\cite{long-lived}.}
In addition, scenarios with $\tau$ neutrinos decaying into three
neutrinos are also dangerous, since, in the absence of fine-tuning,
such models also predict large flavor violating $\tau$ decays (like
$\tau \to 3e$) by SU(2)$_L$ gauge symmetry~\cite{BBN&neut1}.  In
particular, in the cosmological models to be described below, the
resulting flavor-violating $\tau$ decay rates are already excluded by
current bounds. This is because, for these three-body decays, the
flavor-violating $\tau$ branching fraction is of order $(m_\tau
/m_{\nu_\tau})^5\tau_{\nu_\tau}^{-1}$ (to be compared with $(m_\tau
/m_{\nu_\tau})^3\tau_{\nu_\tau}^{-1}$ for the two-body familon
decays), and hence is extremely enhanced for scenarios with neutrino
masses in the currently allowed range.

Therefore, if we adopt a massive unstable neutrino as a solution to
cosmological problems, massless bosons are good candidates for its
decay products. The familon is an example of such a massless boson.
(In the literature, a particular example of a familon, the Majoron, is
often considered.)  If neutrinos decay to familons, cosmological
scenarios, which require specific ranges for neutrino masses and
lifetimes, then predict rates for familon signals in future
experiments, or may even be excluded from current familon bounds,
assuming an absence of fine-tuning. In this section, we first review
some of the potentially interesting cosmological scenarios that
require massive neutrinos.  Then, assuming that the massive tau
neutrino decays through $\nu_\tau\to\nu_l  f$, where $l = e, \mu$,
we will discuss how well these scenarios may be constrained by current
and future collider searches for familons.

Among the several cosmological motivations for massive neutrinos is
the ``crisis'' in the standard BBN scenario. The standard BBN scenario
contains only one free parameter, the baryon to photon ratio $\eta$;
the abundances of the light elements are predicted once we fix
$\eta$. Until a few years ago, the theoretical prediction with
$\eta\sim 3\times 10^{-10}$ was in good agreement with observations.
Recently, however, it has been claimed that the predictions of the
standard BBN are disfavored by observations of the light element
abundances~\cite{BBN_crisis1,BBN_crisis2}: normalizing $\eta$ with the
D and $^3$He abundances, the observed $^4$He abundance is claimed to
be smaller than the standard BBN prediction. There are several
arguments against this viewpoint on the observational side. For
example, the apparent discrepancy vanishes if one adopts a larger
systematic error in the observed $^4$He
abundance~\cite{BBN_crisis2,G_Fuller}, or if the recently measured D
abundance in high red-shift QSO absorber systems is regarded as a
primordial one~\cite{PRD55-540}.\footnote{One should note that there
is another measurement of D abundances that conflicts with the one
preferred by the standard BBN scenario~\cite{PRD55-540}. Thus, this
issue is still an open question.}

On the other hand, if we regard this ``crisis'' as a genuine problem
with the standard BBN theory, it can be taken as an indication of new
physics beyond the standard model. There are several attempts to solve
this crisis by a modification of the standard
scenario~\cite{PRL77-3712,aph9612237,aph9705148}. Here, we
concentrate on a solution that uses massive unstable neutrinos to
reduce the predicted $^4$He abundance. Since the $^4$He abundance
decreases as the energy density at the neutron freeze out time
decreases, the $^4$He abundance becomes smaller if $N_{\nu}$, the
``effective number of neutrino species'' at the neutron freeze out
time, is reduced.  In the standard BBN, $N_{\nu}$ is 3, but it can be
smaller if heavy neutrinos decay and effectively convert their energy
density into lighter particles.  For example, in the presence of the
decay mode $\nu_\tau\to\nu_l  f$, if all the tau neutrinos are
converted into thermal familons and light neutrinos, $N_\nu\simeq
2.6$.  BBN scenarios with massive neutrinos decaying to familons have
been discussed in
Refs.~\cite{BBN&neut1,BBN&neut2,aph9612085,aph9705148}.
The most recent calculation shows that a massive neutrino with
$m_{\nu_\tau}\sim 10-20$ MeV and $\tau_{\nu_\tau}\sim 10^{-2}-1$ sec
can resolve the conflict between theory and
observation~\cite{aph9705148}.

Decaying massive neutrinos are also interesting for large scale
structure formation. The standard cold dark matter (CDM) scenario,
which assumes a flat universe, a scale-invariant initial spectrum, and
that the universe is mostly filled with slowly moving (``cold'')
particles~\cite{CDM}, is very
attractive in explaining the origin of large scale structure. However,
if the normalization of the power spectrum is fixed by the anisotropy
in the temperature of the cosmic background radiation observed by
COBE, the standard CDM scenario predicts too large density
fluctuations at small scales ($\lambda\lesssim 100$ Mpc). Attempts to
explain the scale dependence of the density perturbations include
proposals of CDM with a small component of hot dark matter or with a
cosmological constant~\cite{aph9707285}, or scenarios with a
tilted initial density fluctuation~\cite{tilt}.

As pointed out in Refs.~\cite{prl72-3754,prd51-2669,aph9707143},
CDM with late decaying neutrinos also provides a solution to this
problem. If the neutrino lifetime is long enough, neutrinos dominate
the energy density of the universe at the temperature $T\sim
m_{\nu_\tau}$.  After this stage, the mass density of the neutrino,
$\rho_\nu$, scales as $T^3$. Neutrinos then decay at time $t\sim
(\rho_\nu^{1/2}/M_{\rm pl})^{-1}\sim \tau_{\nu_\tau}$. Once they
decay, the energy density of the neutrinos is converted to radiation
energy density, resulting in an increase of the radiation energy
density without affecting the background photons. This then delays the
time of matter-radiation equality, the matter-dominated era starts
later, and, with the COBE normalization, the density perturbations at
small scales are reduced.  Due to the neutrino decay, the energy
density of the radiation is increased by the factor
$\sim\rho_\nu(T_{\rm D})/T_{\rm D}^4\sim
(m_{\nu_\tau}^2\tau_{\nu_\tau})^{2/3}/M_{\rm pl}^{2/3}$, where $T_{\rm
D}$ is the temperature just before the neutrino decay.  As we can see,
physics (approximately) depends on the combination
$m_{\nu_\tau}^2\tau_{\nu_\tau}$. To obtain the correct density
fluctuation at small scales, this combination must lie in the range
$(m_{\nu_\tau}/{\rm keV})^2(\tau_{\nu_\tau}/{\rm yr})\sim
50-150$~\cite{aph9707143}. 

In these scenarios, the neutrino mass must also lie in a specific
interval. The neutrino must be heavier than about 50~eV; otherwise,
its mass density is always smaller than (or at most comparable to) the
mass density of the CDM, and the scenario does not work well. On the
other hand, if the neutrino mass is above $\sim$ 1 MeV, the neutrinos
decouple from the thermal bath after becoming non-relativistic, and
their number density is reduced.  In this case, the constraint given
above is not applicable.  Furthermore, if the neutrino mass is in the
range $\sim$ 1 -- 10 MeV while the lifetime is longer than $\sim$ 1
sec, the neutrino mass density may be so large at the neutron freeze
out time that $^4$He can be overproduced. Such lifetimes and large
neutrino masses are therefore also disfavored from BBN
considerations~\cite{aph9612085}. In this section, we consider
the mass range $50\text{ eV}\lesssim m_{\nu_\tau}\lesssim 10\text{
MeV}$, with the above caveats in mind.

To summarize, we consider the following two cosmological
scenarios:
\begin{itemize}
  \item BBN
        : $m_{\nu_\tau}\sim 10-20$ MeV,
        and $\tau_{\nu_\tau}\sim 10^{-2}-1$ sec.
  \item Structure formation :
        $(m_{\nu_\tau}/\text{keV})^2(\tau_{\nu_\tau}/\text{yr})
        \sim 50-150$, with
        $50\text{ eV}\lesssim m_{\nu_\tau}\lesssim 10\text{ MeV}$.
\end{itemize}
These scenarios require decays to familons
$\nu_\tau\to\nu_l f$. As we discussed previously, this process is
related to the decay modes $\tau\to l f$ and $b\to s f$
though SU(2)$_L$ and GUT gauge symmetries, respectively.  Thus,
searches for these decay modes are interesting tests of these
scenarios.

\begin{figure}[t]
\centerline{\epsfxsize=0.6\textwidth \epsfbox{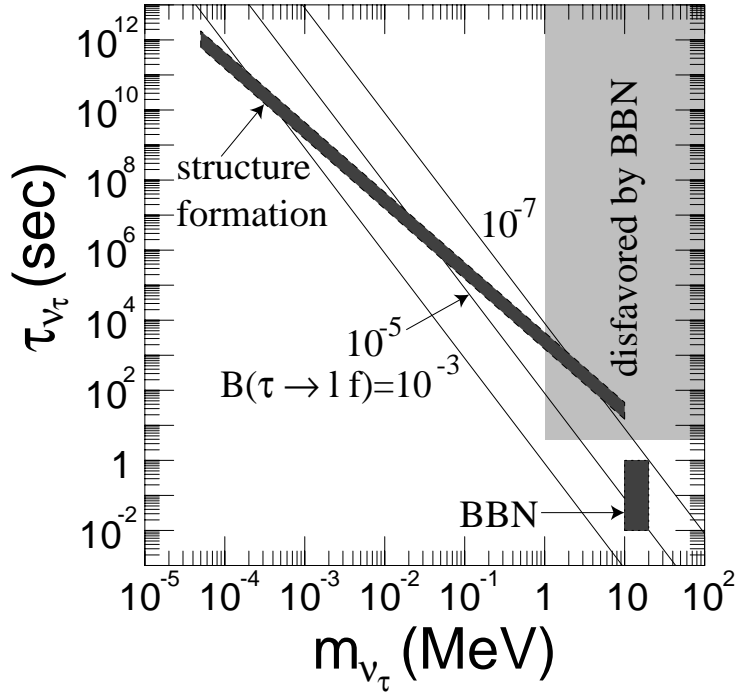}}
\caption{Contours of constant $B(\tau\to l f)$ ($10^{-3}$,
$10^{-5}$, and $10^{-7}$, from below), assuming
$F^L_{\nu_\tau\nu_{l}}=F^L_{\tau l}$, the natural SU(2)$_L$ gauge
relation in the absence of fine-tuning. The regions preferred by the
BBN and structure formation scenarios discussed in the text are also
shown. Note that the parameter region for structure formation with
$m_{\nu_\tau}\protect\gtrsim 1$ MeV may be unreliable because of
limitations in the approximations used to derive the preferred region.
The lightly shaded region is disfavored by BBN.}
\label{fig:mvst_tau}
\end{figure}

Let us start with the $\tau$ familon decay mode. If the decay mode
$\nu_\tau\to\nu_l f$ exists, by SU(2)$_L$ symmetry, the charged
$\tau$ lepton must also have flavor-changing couplings to familons:

\begin{eqnarray}
  {\cal L}_{ f} =  
  \frac{1}{F}\partial_\mu f \left( g_L^{\nu_{\tau} \nu_l}
\bar{\nu}_\tau\gamma^\mu P_L \nu_{l}
  + g_L^{\tau l}
\bar{\tau}\gamma^\mu P_L l\right) + \text{ h.c.}
 \label{L_lepton}
\end{eqnarray}
If there is no fine-tuning, $g_L^{\tau l}\approx
g_L^{\nu_\tau\nu_{l}}$.  Notice that, if the right-handed leptons also
transform under the flavor group, they also couple to familons, and
such interactions may increase the rare $\tau$ decay rate.  The
following argument is therefore conservative.  From the above
Lagrangian, we obtain the decay rate

\begin{eqnarray}
  \Gamma (\tau\to l f) = 
  \frac{m_{\tau}^3}{32\pi F^{L\; 2}_{\tau l}} \ ,
 \label{Gamma_lfv}
\end{eqnarray}
and, using Eq.~(\ref{tau_nu}), we find
\begin{eqnarray}
  B(\tau\to l f) &=& 
  \frac{1}{2} 
  \left(\frac{m_{\tau}}{m_{\nu_\tau}}\right)^3
  \left(\frac{\tau_{\nu_\tau}}{\tau_\tau}\right)^{-1}
  \left(\frac{F^L_{\nu_\tau\nu_{l}}}{F^L_{\tau l}}\right)^2
 \nonumber \\
  &\simeq&
  8.1\times 10^{-4} \times
  \left(\frac{{\rm 1~MeV}}{m_{\nu_\tau}}\right)^3
  \left(\frac{\rm 1~sec}{\tau_{\nu_\tau}}\right)
  \left(\frac{F^L_{\nu_\tau\nu_{l}}}{F^L_{\tau l}}\right)^2 \ .
 \label{Gamma_tau}
\end{eqnarray}
In Fig.~\ref{fig:mvst_tau} contours of constant $B(\tau\to l f)$
are shown in the $(m_{\nu_\tau}, \tau_{\nu_\tau})$ plane, assuming
$F^L_{\nu_\tau\nu_{l}}=F^L_{\tau l}$.

{}From Eq.~(\ref{Gamma_tau}), we see that, in the absence of
fine-tuning, the BBN scenario with decaying neutrinos predicts
$B(\tau\to l f)\sim 10^{-4}$ to $10^{-7}$. The current bounds
$B(\tau^- \to \mu^-  f) < 4.6 \times 10^{-3}\text{ (95\% CL)}$ and
$B(\tau^- \to e^-  f) < 2.6 \times 10^{-3}\text{ (95\%
CL)}$~\cite{albrecht} therefore do not constrain this scenario.
However, if the sensitivity of future experiments is improved by one
order of magnitude or more, the predictions of this scenario may be
tested, and, if the scenario is correct, exotic $\tau$ decays may be
seen. (See Fig.~\ref{fig:mvst_tau}.)

The scenarios motivated by structure formation are also interesting.
In this case, Eq.~(\ref{Gamma_tau}) implies $B(\tau\to l f)\sim
10^{-2}$ to $10^{-8}$. Comparing this result with the current bound,
part of the parameter region of this scenario is already excluded.  As
discussed in Sec.~\ref{sec:taubounds}, future experiments may reach a
sensitivity for $B(\tau \to l  f)$ of $10^{-3}$ or possibly
$10^{-4}$. Thus, if the CDM scenario were realized, the familon decay
mode is likely to be found if the neutrino is lighter than $\sim 1-10$
keV.  The region disfavored by BBN~\cite{aph9612085} is also
shown by a light shading.

\begin{figure}[t]
\centerline{\epsfxsize=0.6\textwidth \epsfbox{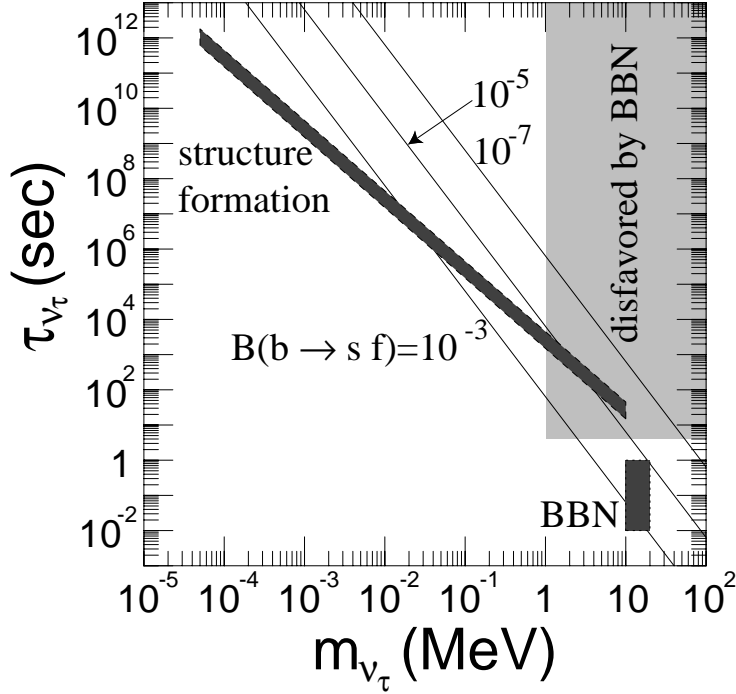}}
\caption{As in Fig.~\protect\ref{fig:mvst_tau}, but with contours of 
constant $B(b\to s f)$ ($10^{-3}$, $10^{-5}$, and $10^{-7}$, from
below), assuming $F^L_{\nu_\tau\nu_{\mu}}\simeq F^R_{bs}$, as would be
the case in GUTs without fine-tuning.}
\label{fig:mvst_bot} 
\end{figure}

Up to now, we have only used SU(2)$_L$ gauge symmetry to relate the
neutrino-familon interaction to existing and future constraints on
$\tau$ decays.  However, if we assume that the same familon also
couples to down-type quarks, the above cosmological scenarios may also
be probed by rare $b$ decays. Such is the case in SU(5) GUTs, where
the lepton doublet and right-handed down-type quarks are in the same
multiplet, and so we also have a coupling of the form
\begin{eqnarray}
  {\cal L} \supset 
  \frac{1}{F} \partial_\mu f g_R^{bs} 
   \bar{b}\gamma^\mu P_R s + \text{ h.c.} \ ,
 \label{L_down}
\end{eqnarray}
and similarly for $d$. With this Lagrangian, we obtain
\begin{eqnarray}
  B(b\to s f) &=& 
  \frac{1}{2}
  \left(\frac{m_b}{m_{\nu_\tau}}\right)^3
  \left(\frac{\tau_{\nu_\tau}}{\tau_b}\right)^{-1}
  \left(\frac{F^L_{\nu_\tau\nu_{l}}}{F^R_{bs}}\right)^2
 \nonumber \\
  &\simeq&
  7.3\times 10^{-2} \times
  \left(\frac{{\rm 1~MeV}}{m_{\nu_\tau}}\right)^3
  \left(\frac{\rm 1~sec}{\tau_{\nu_\tau}}\right)
  \left(\frac{F^L_{\nu_\tau\nu_{l}}}{F^R_{bs}}\right)^2 \ ,
\label{Gamma_b} \end{eqnarray}
where we have used $m_b=4.5$ GeV, and $\tau_b=\tau_B=1.6\times
10^{-12}$ sec. Comparing Eq.~(\ref{Gamma_b}) with
Eq.~(\ref{Gamma_tau}), we can see that the branching ratio $B(b\to
s f)$ is enhanced by about two orders of magnitude relative to
$B(\tau\to l f)$. This results from an enhancement by a factor
$(m_b/m_\tau)^3$, and also the fact that the total decay rate of the
$b$ quark is $V_{cb}$ suppressed, and so is even smaller than that of
the $\tau$ lepton, despite its larger mass.  Contours of constant
$B(b\to s f)$ are shown in Fig.~\ref{fig:mvst_bot}. For the
cosmologically-motivated scenarios, the ranges of $B(b\to s f)$ are
$10^{-2}$ to $10^{-5}$ (BBN), and 1 to $10^{-6}$ (structure
formation).  Because the $b$ decay rates are so enhanced, in each
case, the high value in the predicted range is above bounds which can
be expected from the current data. Future experiments may reach a
sensitivity of $B(b \to s  f) \sim 10^{-4}$ (corresponding to $B(B
\to K  f) \sim 10 ^{-5}$ in Sec.~\ref{sec:bbounds}) and 
will thus cover most of the preferred parameter regions.

Finally we comment on $m_{\nu_\tau}$ measurements at future $B$
factories. As we mentioned earlier, the upper bound on $m_{\nu_\tau}$
will be significantly improved at future $B$ factories, to the level
of 1 -- 3 MeV.  The parameter region that will be covered has a
significant overlap with the neutrino mass required in the above
mentioned cosmological scenarios, and hence such mass measurements
provide another probe of these scenarios. The BBN scenario with
massive unstable neutrinos will be fully tested by the tau neutrino
mass measurement in future $B$ factories. On the other hand, for the
structure formation scenario, most of the interesting parameter region
($m_{\nu_\tau}\lesssim 1$ MeV) may be covered by the search for $b\to
s f$, and even if the neutrino mass is above $\sim$ 1 MeV, this
scenario can be checked by the direct measurement of the tau neutrino
mass (though this region is disfavored by BBN). Therefore, in this
case, measurements of the mass and the branching ratios will have
complementary roles.

\section{Conclusions}
\label{sec:conclusions}

If global family symmetries play a role in determining the patterns of
masses and mixings of the quarks and leptons, they must be
spontaneously broken.  Familons (or Majorons), the massless Goldstone
bosons associated with these broken symmetries, allow rare
opportunities to probe the physics at very high mass scales in a
multitude of low energy settings, and their discovery will signal a
breakthrough in attempts to understand the flavor structure of the
standard model.

The experimental investigation of familons has in the past focussed on
familons coupled to the first two generations.  As we reviewed, such
investigations have led to stringent lower bounds on the flavor
breaking scale $F$ of $\sim 10^{9}$ GeV and $\sim 10^{11}$ GeV in the
leptonic and hadronic sectors, respectively. In contrast, bounds for
familons coupling to the third generation are much less thoroughly
studied.  In the lepton sector, constraints from rare $\tau$ decays
lead to constraints $F \agt 10^6$ GeV; in the hadronic sector, no
bounds have been previously reported.  The lack of study of third
generation familons is conspicuous, especially in light of their
cosmological relevance and the upcoming $B$ factory experiments, which
hold promise for studying $b$ and $\tau$ decays with great precision.

Motivated by these considerations, we have presented a large and
eclectic group of bounds which we believe place the most stringent
constraints on third generation familon couplings.  As emphasized in
Sec.~\ref{sec:collider}, the experimental and astrophysical
implications of familons vary strongly with the underlying flavor
symmetry and depend on whether the flavor symmetry is real or complex,
the familon couplings are axial or vector-like, and whether they are
flavor-diagonal or non-diagonal.  For instance, bounds on $K$ decay in
the presence of mixing effects may in general impose bounds of order
$10^9$ GeV on third-generation flavor breaking scales, but there are
classes of models in which such bounds do not apply.  It is therefore
important to consider a wide variety of experimental signatures.
Different signatures are also related through some well-motivated
theoretical considerations: in the absence of fine-tuning, the familon
couplings of particles related by gauge symmetry are expected to be
similar in strength.  Probes of $\tau$ decays to familons are
therefore indirect probes of $\nu_\tau$ familon couplings, and in
SU(5) grand unified theories, these are both related to $b$ decays.

We began by considering bounds from currently available accelerator
data.  Present values for neutral meson mass splittings imply bounds
on the flavor scale of $\sim 10^5$ to $10^6$ GeV for real familons.
Bounds from neutral meson decays to leptons are significantly weaker,
at the level of $10^3$ GeV, and require both hadronic and leptonic
couplings. More promising are bounds from exotic $b$ decays at LEP.
By extrapolating from current bounds on $b \to s \nu \bar{\nu}$, we
estimate that an analysis of currently available LEP data may provide
a sensitivity to $B(b\to s f)$ at the level of $1.8 \times
10^{-3}$, leading to probes of flavor scales of the order of $10^7$
GeV.

Familons also have astrophysical implications, as they may lead to
anomalously fast cooling of supernovae, red giants, and white dwarfs.
Bounds from direct couplings to $\nu_{\tau}$ are generally weak.
However, couplings of familons to particles of the third generation
may also induce couplings to electrons and nucleons radiatively or
through flavor mixing effects.  Bounds on such couplings are
model-dependent, but may be stringent; in the simple case where a
familon couples diagonally to $t$ quarks, a bound of $F > 10^9$ GeV
from radiatively induced couplings may be set.

Finally, having evaluated a host of new bounds, we considered the
prospects for analyses at future $B$ factories.  Such colliders are
ideal experimental environments for searches for rare $\tau$ and $b$
decay modes and are expected to have greatly improved statistics.  We
find that probes of branching fractions of $10^{-3}$ ($10^{-5}$) for
$\tau$ ($b$) decays may be possible. As discussed in
Sec.~\ref{sec:cosmology}, under the assumption that the flavor scales
for these decays are naturally related to the scales for $\nu_{\tau}$
couplings, such precise probes are sensitive to parameter regions
favored by various BBN and structure formation scenarios, where a
massive unstable neutrino is motivated by possible discrepancies in
cosmological data.  In fact, parts of the parameter regions in such
scenarios are already excluded, and future searches will be able to
explore large portions of the cosmologically-favored parameter space.
Given the present lack of analyses studying third generation familon
couplings, studies at all of these experiments, and particularly
the $B$ factories, are strongly encouraged.

\vspace*{0.1in} 
{\em Note added.}  After the completion of this work,
Ref.~\cite{Hannestad} appeared, in which the impact of decays
$\nu_{\tau} \to \nu_e f$ on BBN was examined.  Tau neutrino masses of
0.1 to 1 MeV and lifetimes of 50 to $2\times10^4$ s were found to be
allowed, implying rates for $\tau \to ef$ and $b \to df$ within reach
of future experiments. (See Figs.~\ref{fig:mvst_tau} and
\ref{fig:mvst_bot}.)

\acknowledgements

We thank L.~Hall and M.~Suzuki for discussions and R.~Patterson,
A.~Seiden, and S.~Stone for valuable correspondence.  This work was
supported in part by the Director, Office of Energy Research, Office
of High Energy and Nuclear Physics, Division of High Energy Physics of
the U.S. Department of Energy under Contract DE--AC03--76SF00098, and
in part by the National Science Foundation under grant PHY--95--14797.
J.L.F. and T.M. thank the Aspen Center for Physics for hospitality
while this work was being completed. J.L.F. is supported in part by a
Miller Institute Research Fellowship.  H.M. is also supported by the
Alfred P. Sloan Foundation. E.S. is supported by a BASF research
fellowship and the Studienstiftung des deutschen Volkes.

\end{document}